\documentclass[
 preprint,
 amsmath,amssymb,
 aps]{revtex4-2}

\usepackage{graphicx}
\usepackage{dcolumn}
\usepackage{bm}
\usepackage{siunitx}
\usepackage{hyperref}
\usepackage{makecell}
\footnotesize

\begin{document}

\title{Automatic generation of input files with optimised k-point meshes for Quantum Espresso self-consistent field single point total energy calculations.}

\author{Elena Patyukova}
 \email{patyukova@gmail.com}
 \altaffiliation{Scientific Computing Department, Daresbury Laboratory, STFC UKRI, UK.}
\author{Junwen Yin}%
\email{junwen.yin@stfc.ac.uk}
 \altaffiliation{Scientific Computing Department, Daresbury Laboratory, STFC UKRI, UK.}
\author{Susmita Basak}%
 \altaffiliation{Scientific Computing Department, Rutherford Appleton Laboratory, STFC UKRI, UK.}
\author{Samuel Pinilla Sanchez}%
 \altaffiliation{Scientific Computing Department, Rutherford Appleton Laboratory, STFC UKRI, UK.}
\author{Alin Elena}%
 \altaffiliation{Scientific Computing Department, Daresbury Laboratory, STFC UKRI, UK.}
\author{Gilberto Teobaldi}%
 \altaffiliation{Scientific Computing Department, Rutherford Appleton Laboratory, STFC UKRI, UK.}

\date{\today}

\begin{abstract}
Performing density functional theory (DFT) calculations requires a careful choice of computational parameters to ensure convergence and obtain meaningful results. This represents a particularly important problem for high-throughput and agentic workflows, where due to computational cost, any additional convergence studies are preferably to be avoided. So, there is a need for tools and models which are able to predict DFT parameters from basic input information, such as a structure. In this work, we develop a machine learning approach to predict the appropriate k-point sampling in DFT calculations and generate the input files for Quantum Espresso self-consistent field calculations. To achieve this, we first generated a training dataset comprising over 20,000 materials, each with an energy convergence threshold of 1 meV/atom. Several ML models were evaluated for their ability to predict k-points distance, and uncertainty estimation was incorporated to guarantee that, for at least 85-95\% of compounds, the predicted k-distance lies within the convergence region. The best-performing models are made publicly available through an open-access web application.

\end{abstract}

\maketitle

\tableofcontents

\section{Introduction}
High-throughput calculations in the age of AI are an invaluable source of data that drives materials discovery pipelines and the development of new models and algorithms. Performing high-throughput DFT calculations requires the appropriate choice of calculation parameters. In this paper, we focus on the most common type of (periodic boundary, plane wave) density functional theory (DFT) calculations — the self-consistent field (SCF) calculation, which involves choosing an appropriate software package (based on material system and targeted properties), selecting a suitable exchange–correlation functional and pseudopotentials, defining a sufficiently dense k-point mesh, applying an appropriate electronic smearing scheme and Gaussian broadening width, and setting suitable plane wave cutoff energies for the wavefunctions and charge density, among other parameters.

Choosing the right parameters is crucial for achieving accurate results while optimizing computational resources, as improper settings can lead to convergence issues or unnecessarily long computation times. However, choosing the optimal strategy of parameter choice is a non-trivial task, as in general, the given level of accuracy in a property can be achieved with multiple sets of parameter combinations; the effects of different parameters are often interdependent and not well understood.

One of the most common approaches to parameter selection in high-throughput calculations is to fix all relevant parameters at certain values for all compounds, usually at sufficiently large k-point densities and energy/density cutoffs. The situation is further complicated by the fact that the rationale behind the chosen fixed parameter values is often not reported. In this situation, there is little control or understanding of the associated errors; calculations often turn out to be partially over-converged and partially under-converged. 

In recent years, there has been great interest and progress in automating high-throughput workflows both through rigid-logic automation systems such as AiiDA~\cite{Huber2020}, Atomate2~\cite{Ganose2025}, Pyiron~\cite{JANSSEN2019}, AFLOW~\cite{Curtarolo2012}, and new emerging agentic LLM-driven frameworks~\cite{Zou2025, wang2025dreamsdensityfunctionaltheory, Liu2025}. Fixed workflow systems provide robust workflows, data management tools, HPC-interfaces, input/output processing and analysis, and help in standardising DFT simulations, results reproducibility, and reuse. Emerging agentic frameworks use more flexible approaches and aim at performing the full cycle of calculations and data analysis in automated manner. Potentially these frameworks can give wide access to theoretical tools to non-experts and robots.  However, the choice of calculation parameters for those workflows remains a problem in both cases.

In this work, we aim to improve on the conventional baseline of fixed parameters for single-point SCF energy calculations conducted by Quantum ESPRESSO~\cite{Giannozzi_2009} by developing a tool using machine learning to  predict optimal calculation parameters. By sharing this tool we aim to help to improve the quality of high-throughput datasets and make calculations greener by reducing unnecessary calculation time.

We begin by reviewing prior work in the field. Based on this analysis and our own convergence studies, we identify an optimal strategy for data generation and subsequently construct the dataset. We then benchmark different models to determine which achieves the lowest prediction errors for k-point density. A key consideration in this task is uncertainty estimation, since the cost of prediction errors is asymmetric: underestimating the required k-point density can lead to inaccurate results, while overestimating it only increases computational expense. To address this, we predict the 0.85, 0.9, and 0.95 quantiles rather than the median (0.5 quantile), ensuring that at least 85\%, 90\%, and 95\% of our predictions are not underestimated. 

\begin{figure}[h]
\centering
  \includegraphics[width=\textwidth]{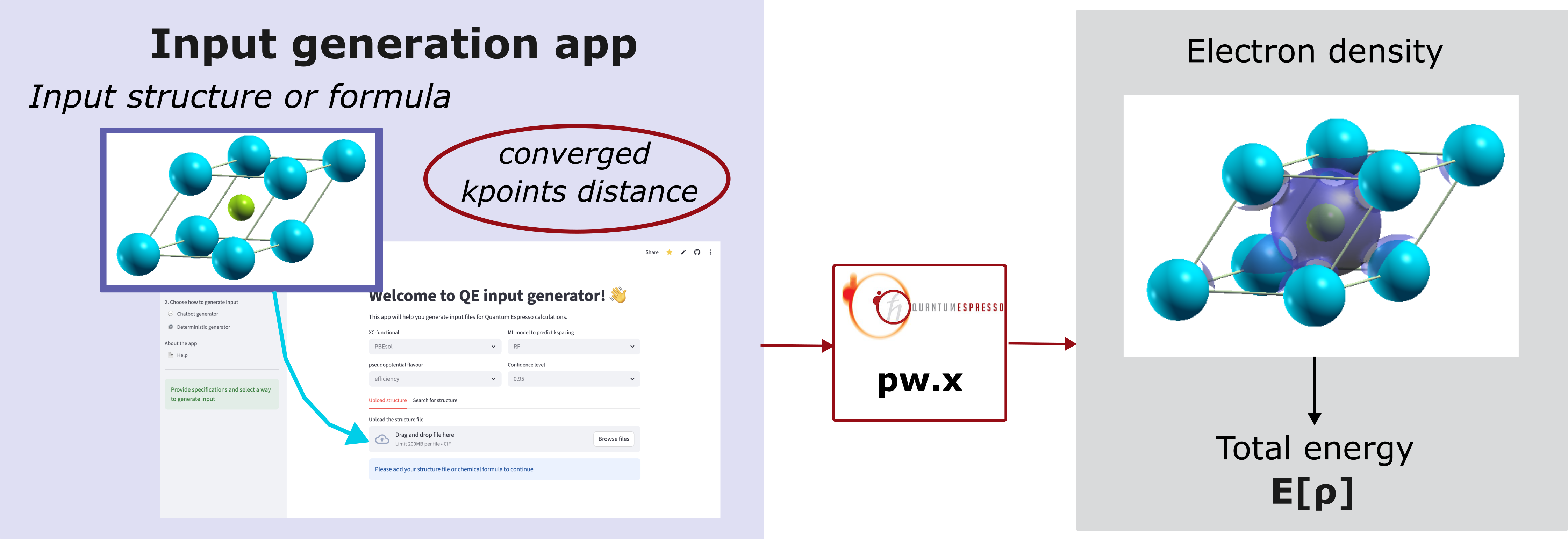}
  \caption{The best model developed during this work is encapsulated in the web application for the generation of input files for QE pw.x calculations.}
  \label{fgr:toc}
\end{figure}

Finally, we encapsulate our findings in a practical web application that automatically generates input files for single-point SCF calculations in Quantum ESPRESSO for any given structure (see Figure~\ref{fgr:toc}). 

\subsection{Contributions of this work}
In this paper, we make the following contributions:
\begin{enumerate}
\item \textbf{A large-scale convergence dataset} of 20,178 compounds, generated with consistent SSSP-1.3 Efficiency pseudopotentials and cold smearing, providing reliable reference k-meshes for SCF single point energy calculation, with convergence criterion at 1 meV/atom.

\item \textbf{A systematic evaluation of machine learning models} (Random Forest, Gradient Boosting, CrabNet, CGCNN, ALIGNN) and feature types (composition, structure, SOAP, lattice descriptors, metallicity embeddings) for predicting optimal k-point densities.

\item \textbf{A compact and interpretable rule-set} derived via surrogate decision-tree modeling, offering intuitive insight into the key physical and chemical factors influencing k-point spacing.

\item \textbf{An uncertainty estimation framework} using conformalised quantile regression (CQR), ensuring that predictions satisfy probabilistic guarantees and avoid underestimating required k-point density.

\item \textbf{A publicly available web application} that automatically generates fully specified Quantum ESPRESSO SCF input files using the trained models

\item \textbf{Quantitative demonstration of computational savings}, showing that ML-driven predictions reduce unnecessary k-mesh density and thereby lower computation time relative to fixed-density baselines.
\end{enumerate}

\subsection{Previous work}
The most common approach to the determination of parameters for the DFT calculation is to perform a convergence curve modeling with respect to different parameters, such as k-point mesh, cutoff energies, and smearing. However, in the case of high-throughput calculations, this is in general unreasonable due to the amount of calculation time required. So, fixing those parameters for all compounds in the high-throughput dataset at some constant chosen by an expert was adopted. There were multiple attempts to improve the quality of this choice.
Choudhary and Tavazza~\cite{Choudhary2019} developed an automatic convergence procedure to derive k-point densities and energy cut-off values to reach a 1 meV/cell convergence in energy for $\sim$ 30,000 materials, using the Vienna Ab initio Simulation Package (VASP). The authors used projector-augmented wave (PAW) pseudopotentials and Gaussian smearing (a fixed value of 0.01 eV) for all calculations. The authors tried to find correlations between k-points, cut-off parameters, and other material properties such as density, number of electrons, the slope of bands, number of band-crossings, the maximum plane-wave cutoff used for pseudo-potential generation, crystal system, the number of unique species in the material, and exchange-correlation functional. They discovered that: (1) convergence is nearly independent of the exchange-correlation functional; (2) converged plane wave cutoff is significantly correlated with the maximum cutoff used in pseudo-potential generation (ENMAX) (Pearson’s correlation $\simeq$ 0.64); (3) Strong correlation exists between the slope of band crossing at Fermi surface for metals and the required k-point density (Pearson’s correlation $\simeq$ 0.66); (4) Certain preferences exist for using higher k-points  (often in cubic and hexagonal) vs higher plane-wave cut-off energies (often in triclinic) for different crystal systems; (5) The presence of some elements, for example, transition metals required on average high k-point densities; presence of electronegative elements (O, F) required high cut-offs. The results imply that the derived convergence data would be expected to depend strongly on the pseudo-potential used, so one cannot use it for calculations with Quantum Espresso and Quantum Espresso’s set of pseudo-potentials.

Janssen et al.~\cite{janssen2021automatedoptimizationconvergenceparameters} conducted a systematic study aiming at understanding the optimal way to choose energy cutoffs and k-point density for DFT calculations with VASP (so smearing was fixed to default VASP values). They considered energy-volume dependence and derived quantities (the bulk modulus in particular as it is more sensitive to fitting/convergence errors)  as the target property to analyse convergence in several elemental metal compounds. They decomposed the total error into a systematic part, obtained by comparing results at finite cutoffs and k-point meshes against highly converged references, and a statistical part, identified as the residual oscillatory noise arising from discrete changes in the number of plane waves or k-points. This separation enabled the construction of error phase diagrams, distinguishing regimes where systematic or statistical errors dominate. Their analysis showed that systematic errors (e.g., from insufficient cutoffs) dominate for most elements, while statistical errors can prevail in transition metals such as Cu, Pd, and Ag. They concluded that in the statistical regime, it is most efficient to refine the cheaper parameter (typically k-points), whereas in the systematic regime, both cutoff and k-point density must be increased simultaneously.

The Quantum ESPRESSO community paid a lot of attention to systematic verification of pseudopotentials~\cite{Prandini2018} and pseudopotential-based codes (the numerical basis sets implemented in these codes including plane waves, Gaussians combined with plane waves, Daubechies wavelets, and atomic orbitals~\cite{Bosoni2024}. They designed a testing protocol (SSSP – Standard Solid-State Pseudopotential protocol) to benchmark pseudopotentials in Quantum ESPRESSO across multiple tasks (equations of state, convergence of phonon frequencies, cohesive energies, pressures, and band structures with respect to plane-wave cutoffs). They tested available pseudopotentials on elemental crystals (while rare-earth nitrides and SiF4 being the exceptions). These investigations resulted in the list of recommended pseudopotentials (including ultrasoft, norm-conserving, and PAW) for each element in the periodic table and recommended energy/density cut-off pair for these pseudopotentials. The authors also pointed out that the required cut-off depends strongly on the property of interest. They also noticed independence of convergence patterns for local/semi-local functionals (and for hybrid/meta-GGA, no dedicated pseudopotentials existed).

Nascimento et al.~\cite{nascimento2025accurateefficientprotocolshighthroughput} extended this line of work by analysing the interplay between k-point sampling and electronic smearing in Quantum ESPRESSO calculations. Building on the SSSP pseudopotential library—with its recommended plane-wave and charge-density cutoffs—they carried out a systematic study on a dataset of 269 experimentally known compounds (including oxides, metals, and insulators, both 3D and 2D) selected from the Materials Cloud databases, with unit cells containing up to 14 atoms. Their goal was to determine how different choices of k-point density and smearing affect both numerical accuracy and the fraction of calculations that fail to converge. The guiding principle was that sampling (statistical) errors from insufficient k-points should be suppressed in favour of controlled systematic errors introduced by finite smearing, since the latter are predictable and tunable. Using Marzari–Vanderbilt cold smearing, they identified three practical protocols—fast, balanced, and stringent—that offer trade-offs between efficiency and precision. The balanced protocol (smearing temperature $\sigma$ = 0.02 Ry, k-point spacing $\lambda$ = \SI{0.15}{\angstrom}$^{-1}$) was shown to work well for most systems, while the stringent protocol ($\sigma$ = 0.0125 Ry, $\lambda$ = \SI{0.1}{\angstrom}$^{-1}$) is necessary for challenging cases such as lanthanides and actinides. In this way, they provided ready-to-use parameter sets that minimize errors and computational cost while ensuring robust convergence across large, chemically diverse datasets.

In the past year, there has been rapid progress in LLM-based multi-agent systems designed for automated computational chemistry workflows, such as El Agente~\cite{Zou2025}, DREAMS (Density Functional Theory Based Research Engine for Agentic Materials Simulation)~\cite{wang2025dreamsdensityfunctionaltheory}, and VASPilot~\cite{Liu2025}. These agents typically employ multiple child agents to handle automated convergence schemes, either by leveraging existing recommendations from the literature and packages such as pymatgen or ASE, or by generating a series of calculations with systematically varied parameters for convergence testing. They offer clear advantages in managing complex tasks—such as input file generation, job submission, error handling, and post-processing—without human intervention. However, a persistent challenge remains in determining the optimal set of calculation parameters that ensures accurate results while minimizing computational cost.

In summary, previous studies have progressively clarified the roles of cutoffs, k-point density, and smearing in determining the accuracy and efficiency of plane-wave DFT calculations. Nevertheless, the optimal choice of these parameters for high-throughput studies remains an open problem. In this work, we continue our work in this direction.

\section{Data Generation and Analysis}
\subsection{Data generation}
With all interconnected factors affecting k-points convergence for single point energy calculations in mind, the following set of parameters was chosen to generate the data:

\begin{enumerate}
\item The valence electrons are represented using the Standard Solid-State Pseudopotential (SSSP) library (version 1.3, PBEsol, efficiency set), with plane-wave cutoff energies determined following the efficiency-level recommendations.

\item The Marzari–Vanderbilt cold-smearing scheme, with a smearing width of 0.01 Ry, is applied in all calculations. The choice of the cold smearing is justified by its ability to yield a free energy that is quite insensitive to variations in the smearing temperature. The smearing temperature is set as 0.01 Ry, which ensures the smearing convergence for most materials according to Nascimento’s report~\cite{nascimento2025accurateefficientprotocolshighthroughput}. This also implies that even for insulators, the smearing is applied to the electronic states around the Fermi level, rather than fixed occupations.

\item The 20187 structures were selected randomly from the MC3D PBEsol-v1 dataset, and no additional structural relaxation was performed~\cite{huber2025mc3dmaterialscloudcomputational}. 

\item Non-spin-polarized calculations were performed for all materials, and no magnetic configurations were taken into consideration. 

\item The k-mesh convergence is determined as follows: when the energy difference among three consecutive k-point distances becomes smaller than 1 meV per atom, the first point of these three is identified as the converged k-mesh. This procedure does not guarantee that one has the optimal set of parameters, all pockets of the Fermi surface are resolved, or that the total energy of compounds with a band gap below 0.14 eV is resolved correctly. However, it is an easy way to generate the data. 

\item We employ the definition of \textbf{k-distance}, as implemented in the AiiDA–QuantumESPRESSO package~\cite{Huber2020,Uhrin2021}, to generate various k-point meshes. The k-distance represents the maximum spacing (in \SI{}{\angstrom}$^{-1}$) between adjacent k-points in reciprocal space, such that the number of k-points along each reciprocal lattice vector $b_i$ is given by $\lceil |b_i| / k_{dist} \rceil$. Starting from a k-distance of \SI{1.0}{\angstrom}$^{-1}$, we systematically scan all possible k-meshes by varying the k-distance in steps of \SI{0.005}{\angstrom}$^{-1}$. We always include the Gamma point for future calculations regarding electronic structures, phonon properties, etc.

\end{enumerate}

The generated data was processed to shape the dataset. This included reducing all systems to primitive cells and then removing all structural duplicates. To do this, we used pymatgen~\cite{Ong2013}. The property which we predict (k-points distance) is intensive and should not depend on the size of the number of formula units, or centering of the cell. Pre-processing ensures that the model predictions are consistent with this physical constraint. The number of k-points that enter the Quantum Espresso input file is easily generated for any unit cell/ super cell from the k-point distance. 

\subsection{Data Analysis}
\begin{figure}[h]
\centering
  \includegraphics[width=\columnwidth]{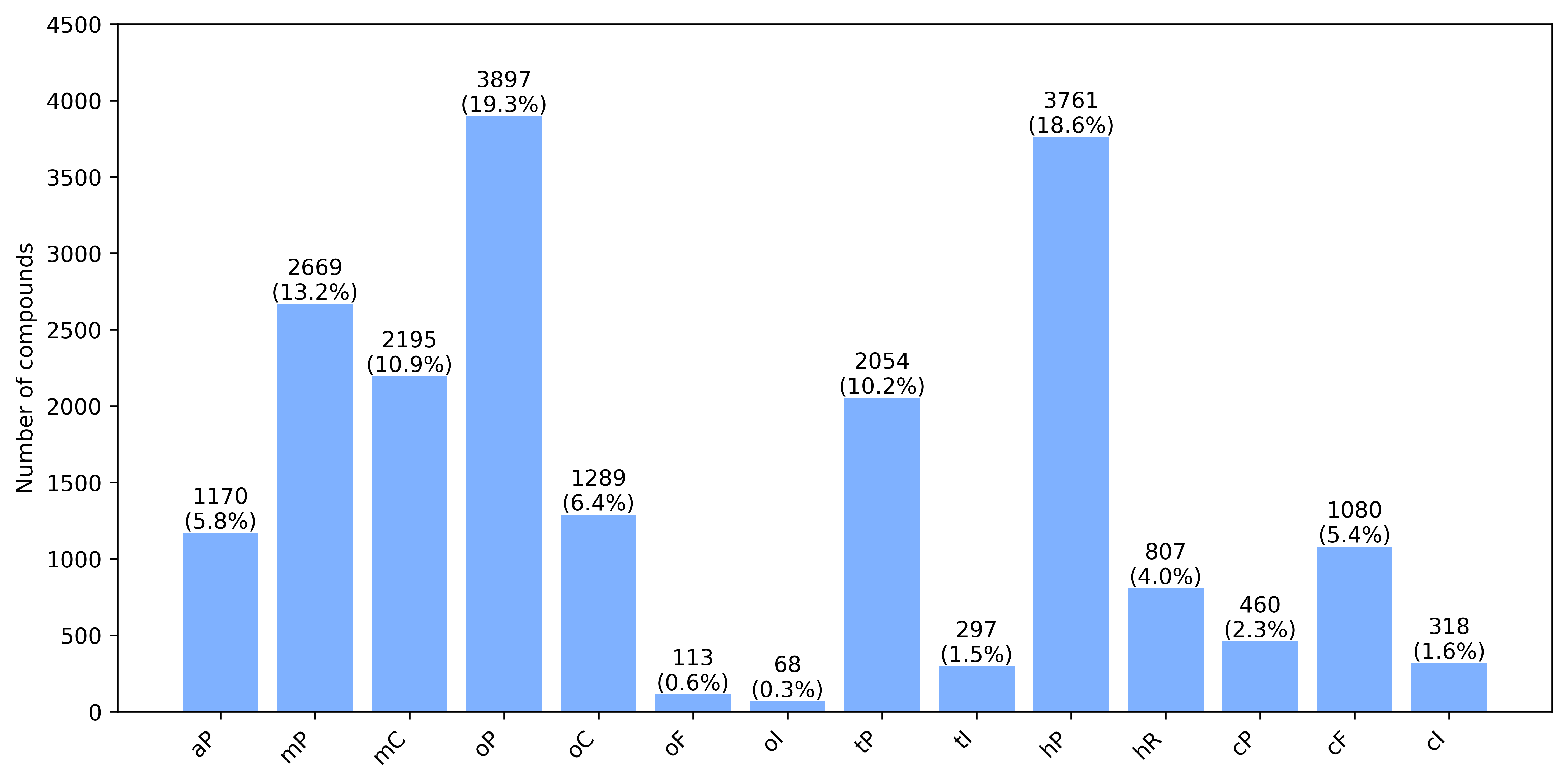}
  \caption{Distribution of compounds in the dataset over 14 Bravais lattices}
  \label{fgr:distr-lattices}
\end{figure}

\begin{figure}[h]
\centering
  \includegraphics[width=\columnwidth]{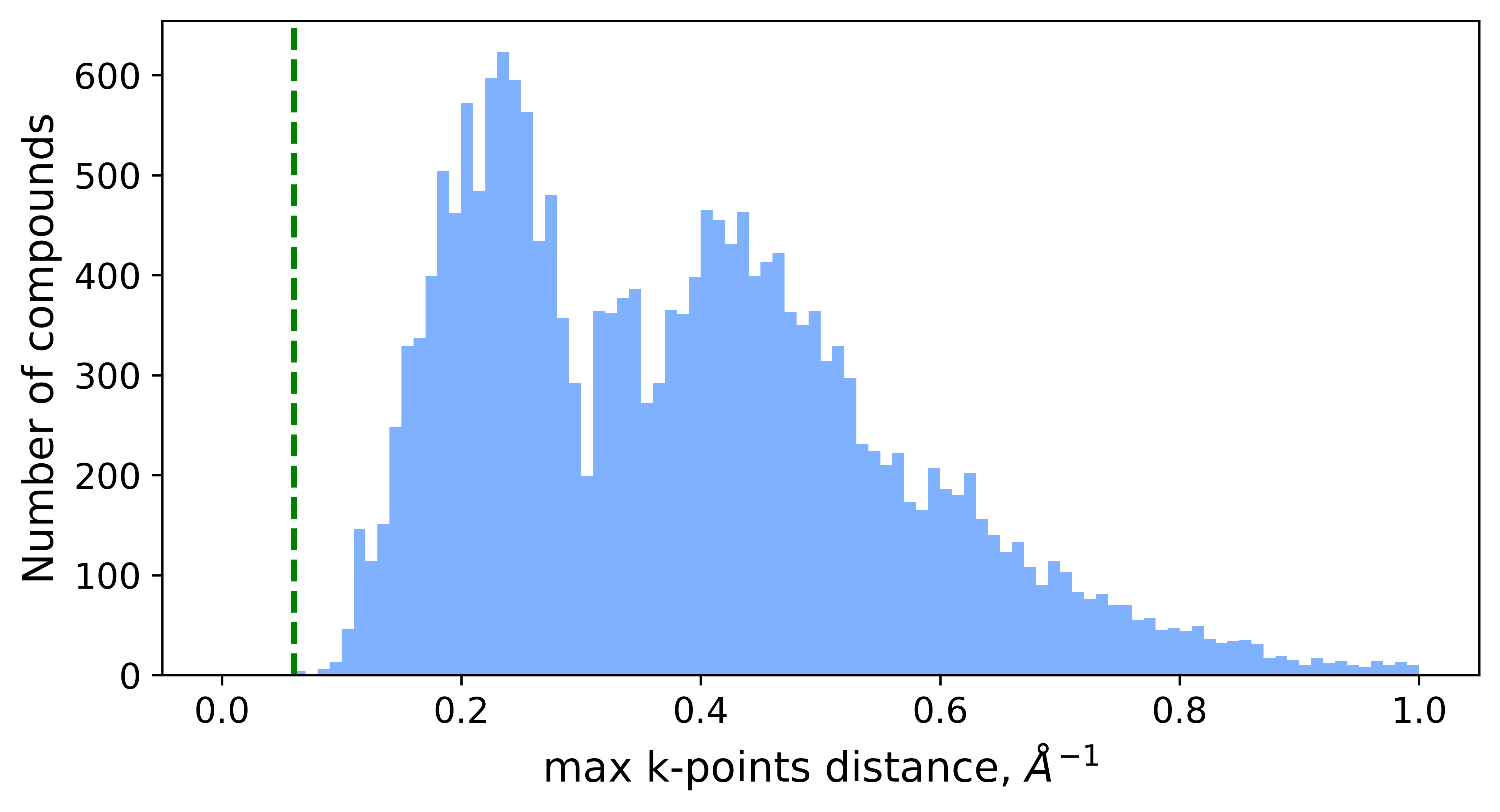}
  \caption{Distribution of compounds with respect to the value of maximum distance between k-points. Vertical dashed line represents a max k-points distance, \SI{0.06}{\angstrom}$^{-1}$), recommended for the reference calculations in [12].}
  \label{fgr:distr-k-distances}
\end{figure}

\begin{figure}[h]
\centering
  \includegraphics[width=\columnwidth]{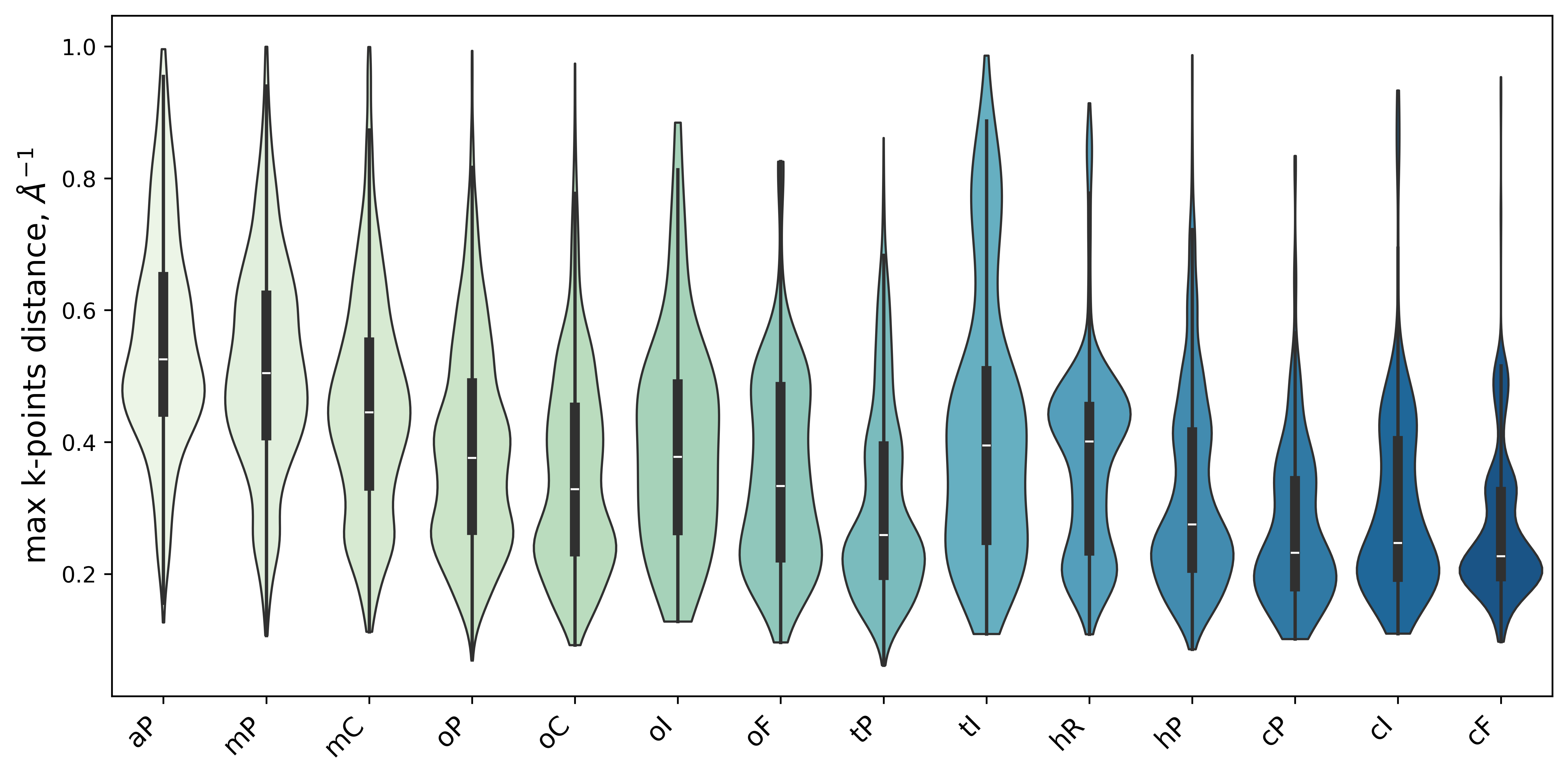}
  \caption{Separate distributions with respect to the maximum distance between k-points for compounds with fixed Bravais lattice.}
  \label{fgr:violin-plot}
\end{figure}

\begin{figure}[h]
\centering
  \includegraphics[width=\columnwidth]{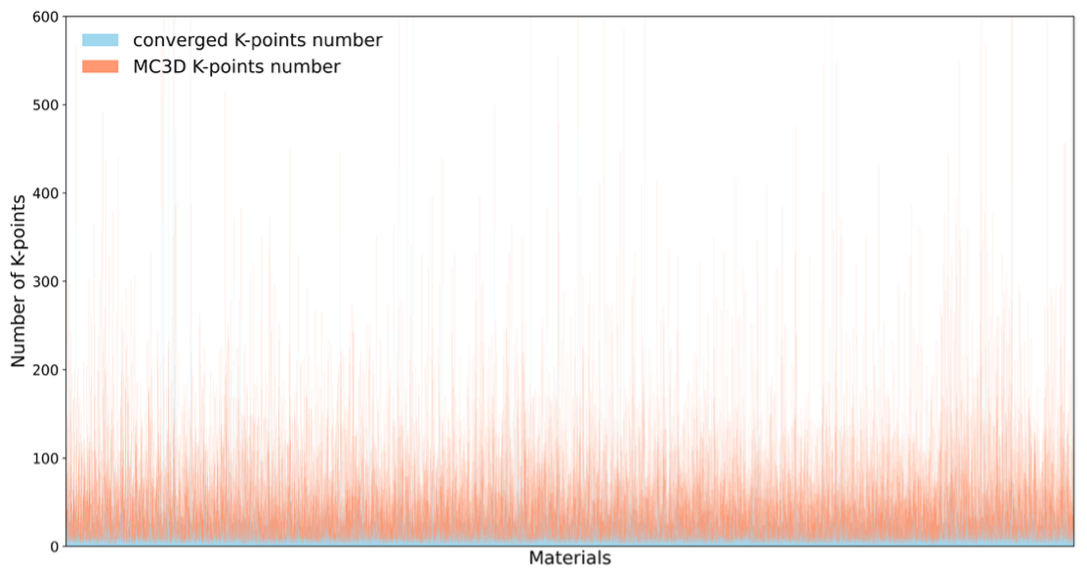}
  \caption{A comparison of the symmetrized K-points numbers used for the same material in MC3D-PBEsol-v1 and the optimized converged calculations in our database.}
  \label{fgr:MC3D-goldilocks}
\end{figure}

The final dataset comprises 20,178 unique crystal structures, encompassing all non-radioactive and some radioactive elements (see element occurrence frequencies in the Supporting Information, SI Figure S1). The dataset exhibits a broad distribution in the number of atoms per primitive unit cell (SI Figure S2) and includes compounds containing up to seven distinct elements (SI Figure S3). Among the 230 crystallographic space groups, 200 are represented. Since the shape of the first Brillouin zone is determined by the Bravais lattice type, it is noteworthy that all 14 Bravais lattices are represented in the dataset (Figure~\ref{fgr:distr-lattices}), ensuring comprehensive coverage of symmetry types relevant for k-point sampling. 
Figure~\ref{fgr:distr-k-distances} shows the distribution of compounds with respect to the maximum distance between k-points, the property predicted by our model. The vertical dashed line indicates the recommended reference value of \SI{0.06}{\angstrom}$^{-1}$ from “How to verify the precision of density-functional-theory…” ~\cite{Bosoni2024}, which aligns well with our dataset, as most of the structures lie just to the right of this threshold. The overall distribution exhibits two distinct peaks, corresponding to metallic systems (lower maximum k-point distance) and insulators (higher distance). Figure~\ref{fgr:violin-plot} presents analogous distributions for subsets of compounds constrained to individual Bravais lattices. There is a clear trend that more symmetric lattices, on average, require dense k-point meshes. So, we can conclude that both metallicity and lattice symmetry significantly influence the optimal k-point density.
To illustrate the efficiency of using an optimized k-mesh relative to commonly used databases, we compare it with several widely adopted materials datasets. For example, the MC3D PBEsol-v1 dataset~\cite{huber2025mc3dmaterialscloudcomputational} employs a default k-point distance of \SI{0.15}{\angstrom}$^{-1}$, which is substantially denser (i.e., over-converged) for approximately 98\% of the materials in our generated database. The Materials Project~\cite{Jain2013} uses a default k-point density of 1000/(number of atoms in the unit cell), which likewise results in a denser-than-necessary k-sampling for about 70\% of our materials. Assuming identical settings for pseudopotentials, smearing, and symmetry reduction (as in MC3D PBEsol-v1), an optimized, convergence-verified k-mesh would enable significant reductions in symmetrized k-point counts—and therefore in overall computational energy consumption, as show in Figure~\ref{fgr:MC3D-goldilocks}.
To evaluate the robustness of our non–spin-polarized calculations, we selected 268 materials evenly sampled across different categories, including metallic and semiconducting systems, the 14 Bravais lattice types, and four ranges of atomic counts (0–10, 10–20, 20–40, and > 40 atoms per cell) for spin-polarized tests. The results show that 82\% of these materials converge to the same k-mesh, while the remaining 18\% exhibit magnetic behaviours in the spin-polarized calculations. As we are unable to capture the magnetic behaviours in non-spin-polarized calculations,  it is understandable that the converged k-mesh is different. We also notice that among all samples, 94\% have denser k-meshes in the non–spin-polarized calculations than those in the spin-polarized calculations, while 6\% have slightly lower densities, indicating that the non–spin-polarized calculations generally provide well-converged results. 
\begin{figure}[h]
\centering
  \includegraphics[width=0.6\columnwidth]{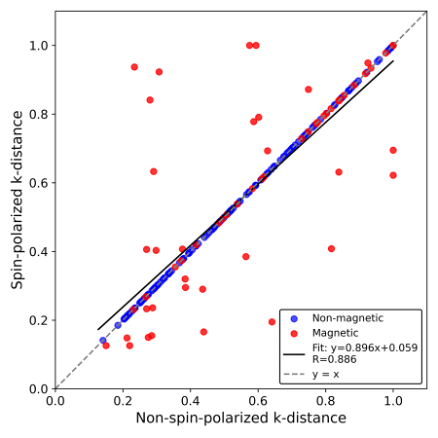}
  \caption{Correlation between converged k-distance in non-spin-polarized and spin-polarized  calculations.}
  \label{fgr:spin-polarised}
\end{figure}
The SSSP Efficiency library (version 1.3) was employed in our training data generation, as it is recommended for high-throughput calculations with affordable computational cost~\cite{Prandini2018}. We acknowledge that, in practical applications, users may prefer the SSSP Precision library to achieve results closer to all-electron calculations. This naturally raises the question of whether our dataset, trained with the Efficiency library, remains valid for predicting k-mesh settings when using the Precision library. To examine this, we selected 215 materials and performed non–spin-polarized calculations using the Precision pseudopotentials. The results show that in 90\% of the cases, the Precision library yields a k-mesh that is either denser than or identical (78\%) to that obtained with the Efficiency library. In the remaining 10\%, 8\% of the total materials show energy differences smaller than 2 meV/atom compared to the converged results, confirming the robustness of the Efficiency-trained dataset.
\begin{figure}[h]
\centering
  \includegraphics[width=0.6\columnwidth]{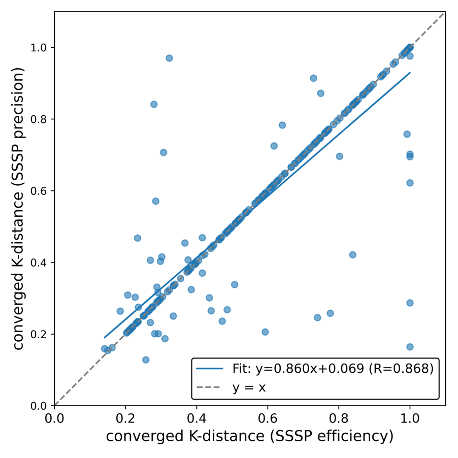}
  \caption{Correlation between converged k-distance in SSSP-efficiency and SSSP-precision calculations.}
  \label{fgr:spin-polarised}
\end{figure}
In the discussions above, we identified the main limitation of our current non–spin-polarized database—its inability to correctly describe magnetic materials. However, determining whether a material is magnetic or non-magnetic is often non-trivial. To provide an initial assessment of this issue, we used the MC3D database as a reference to identify approximately 2,800 potentially magnetic materials in our dataset and performed spin-polarized calculations on them with finite initial magnetic moments. The initial magnetic moments were assigned based on the presence of d or f electrons: materials containing such elements were initialized with their maximum possible magnetic moments (i.e., 5 or 7 $\mu$B for d or f shells, respectively), while all others were assigned a small default value of 0.1 $\mu$B. We found that only about 43\% of the non–spin-polarized calculations yield the same converged k-mesh as their corresponding spin-polarized counterparts for these magnetic materials, indicating that for magnetic materials, spin polarization can significantly affect the convergence behaviour and, consequently, the accuracy of the derived properties.

\section{Developing prediction models}
\subsection{Models to predict k-points density}
To predict the density of k-points, we train several models on our dataset: Random Forest Regression, Gradient Boosting Regression, CrabNet, CGCNN, ALIGNN.

Random Forest Regression~\cite{Breiman2001} is a machine learning method that uses an ensemble of multiple (i.e. 100) decision trees, each trained on a bootstrap sample of the data with a random subset of features. The final prediction is obtained by averaging the predictions of individual trees, which reduces variance and overfitting. We use Random Forest Regressor implemented in scikit-learn [2] library.

Gradient Boosting Regression~\cite{Friedman2001} is another ensemble-based method, which, instead of averaging the predictions of multiple weak learners, builds them sequentially, where each subsequent tree attempts to correct the errors of the previous one. We use the GradientBoostingRegressor implemented in scikit-learn.

Random Forest Regressor and Gradient Boosting Regressor require features in a tabular format representing each compound as a vector derived from the compound’s compositions or/and structure. We use the following sets of features: 
\begin{enumerate}
    \item C = composition features. These are essentially the Magpie~\cite{Ward2016} style features computed with the matminer library ~\cite{Ward2018}. We include ElementProperty, Stociometry, ValenceOrbital. ElementProperty features are composed of statistics ["minimum", "maximum", "range", "mean", "avg\_dev", "mode"] of properties of atoms in the structure. The examples of properties are number, Mendeleev number, atomic weight, electronegativity, etc. Stoichiometric features depend only on the fractions of elements; these include the number of elements in the compound and several $L^p$ norms of the fractions. ValenceOrbital features represent the average number and/or fraction of valence electrons in specified orbitals. 
    \item S = structural features. For the first part of these features, we use GlobalSymmetryFeatures and DensityFeatures from matminer. These include space group number, crystal system, centrosymmetry, number of symmetry operations, density, volume per atom, and packing fraction. For the second part of these feature vectors, we use modified Smooth Overlap of Atomic Positions (SOAP) fingerprints~\cite{Sandip2016}.  The modification consists of: (1) substituting all atoms with ‘X’ atom type; (2) calculating SOAP fingerprints for the structure with anonymised atoms with the DScribe library~\cite{Himanen2020}; (3) averaging resulting SOAP descriptors over all atoms. Modification is required to ensure that features have the same size for all compounds.
    \item L = lattice features. Lattice features include a, b, c, $\alpha$, $\beta$, $\gamma$ parameters for the unit cell, for the inverse unit cell, and the Bravais lattice.
    \item JarvisCFID features with were originally suggested for the prediction of k-points length by Choudhary and Tavazza~\cite{Choudhary2019} and are available in a matminer library~\cite{Ward2018}. These are complex feature vectors with length 1557, which include information about composition, unit cell, core charges, angular distribution function, radial distribution function, dihedral angle distribution function, and nearest neighbours.
    \item M = metallicity features. It is expected that the required k-point density strongly depends on whether a compound is metallic or insulating. In metals, a denser k-point mesh is required to accurately resolve the Fermi surface, whereas in semiconductors and insulators, the Brillouin zone integration converges more rapidly. The reason is that metallic systems have a discontinuity in electronic occupation at the Fermi level, making the total energy highly sensitive to how finely the electronic states are sampled. If the mesh is too coarse, errors in total energy can reach tens of meV per atom, leading to unreliable stability comparisons and inaccuracies in derived properties such as formation energies or phonon spectra. Since metallicity is not known a priori for arbitrary compounds, we train a dedicated CGCNN model on Materials Project data  was to classify materials as metals or non-metals. Performing inference with a trained model for a new compound and extracting learned embeddings, we get numerical vectors which encode information about metallicity for each particular compound, and can be used as additional features in k-points prediction models. So, M features are those internal embeddings from the pre-trained CGCNN model. (The 'is\_metal' dataset was downloaded from MP in July 2025, contained $\sim 180000$ unique structures. Performance of trained CGCNN model on test set is:  accuracy = 0.84, F1\_score = 0.83, Matthews Correlation Coefficient = 0.69. Model checkpoint can be found in code repository.)
\end{enumerate}

CrabNet~\cite{Wang2021} is a composition-based deep learning model that adapts the Transformer architecture~\cite{vaswani2023attentionneed}, originally developed for natural language processing, to materials science applications. In CrabNet, the elements in a chemical formula serve as tokens, analogous to words in a sentence. Each element is associated with a learned embedding vector, while stoichiometric fractions are incorporated through a modified positional-encoding scheme that enables the model to represent not only which elements are present but also their relative proportions. Passing these token–fraction embeddings through stacked self-attention layers allows CrabNet to capture complex inter-element interactions, modelling both pairwise and higher-order chemical relationships.

As input features, we use mat2vec embeddings~\cite{Tshitoyan2019} together with extended cgcnn descriptors, cgcnn+edc. Mat2vec embeddings were developed by Tshitoyan et al.~\cite{Tshitoyan2019} via skip-gram variation of Word2vec method trained on 3.3 million scientific abstracts, and originally used in CrabNet model. CGCNN features were suggested in CGCNN paper~\cite{Xie2018} and elemental attributes such as atomic number, atomic weight, and basic electronic properties. The extended cgcnn feature set (cgcnn + edc) augments the original cgcnn elemental descriptors with information from the SSSP 1.3 pseudopotential library (efficiency, PBEsol), including recommended energy and density cutoffs, and the pseudopotential type (norm-conserving, ultrasoft, etc.). These quantities are element-specific because they depend on the corresponding pseudopotential. Their inclusion is motivated by the expectation that k-point density convergence is intrinsically linked to the chosen energy and density cutoffs~\cite{Choudhary2019}. In a style similar to the cgcnn features, we encode the numbers for cut-off values as one-hot feature vectors of a size suitable for the range of features. Finally, to reduce redundancy in cgcnn+edc descriptors, we perform PCA on them and remove dimensions with negligible information content. 

The next model we try is CGCNN~\cite{Xie2018}, or the Crystal Graph Convolutional Neural Network, which is a graph neural network designed to represent crystalline materials as atom graphs. Here, we consider two ways to build an atom graph: (1) radius graph, which assumes that two atoms are connected with an edge if the distance between them (or their periodic images) is smaller than a cutoff radius. Such a graph usually is a multigraph; (2) Voronoi tessellation graph, which assumes that two atoms are connected by an edge if they share the face of the Voronoi polyhedra. Once the atom graph is constructed, CGCNN employs a gated graph convolution operation for message passing, allowing atom embeddings to be updated based on their local chemical environments. By stacking multiple convolutional layers, the model captures both short- and longer-range structural information, making it effective for predicting a wide variety of materials properties directly from crystal structures.

ALIGNN~\cite{Choudhary2021} conceptually follows the CGCNN model and extends this framework by introducing a line graph in addition to the atom graph, thereby explicitly encoding angular information between neighbouring bonds. This enriched representation enables the model to distinguish the graphs that would be indistinguishable using only information about the atoms’ identities and distances between them. 

\begin{figure}[h]
\centering
  \includegraphics[width=\columnwidth]{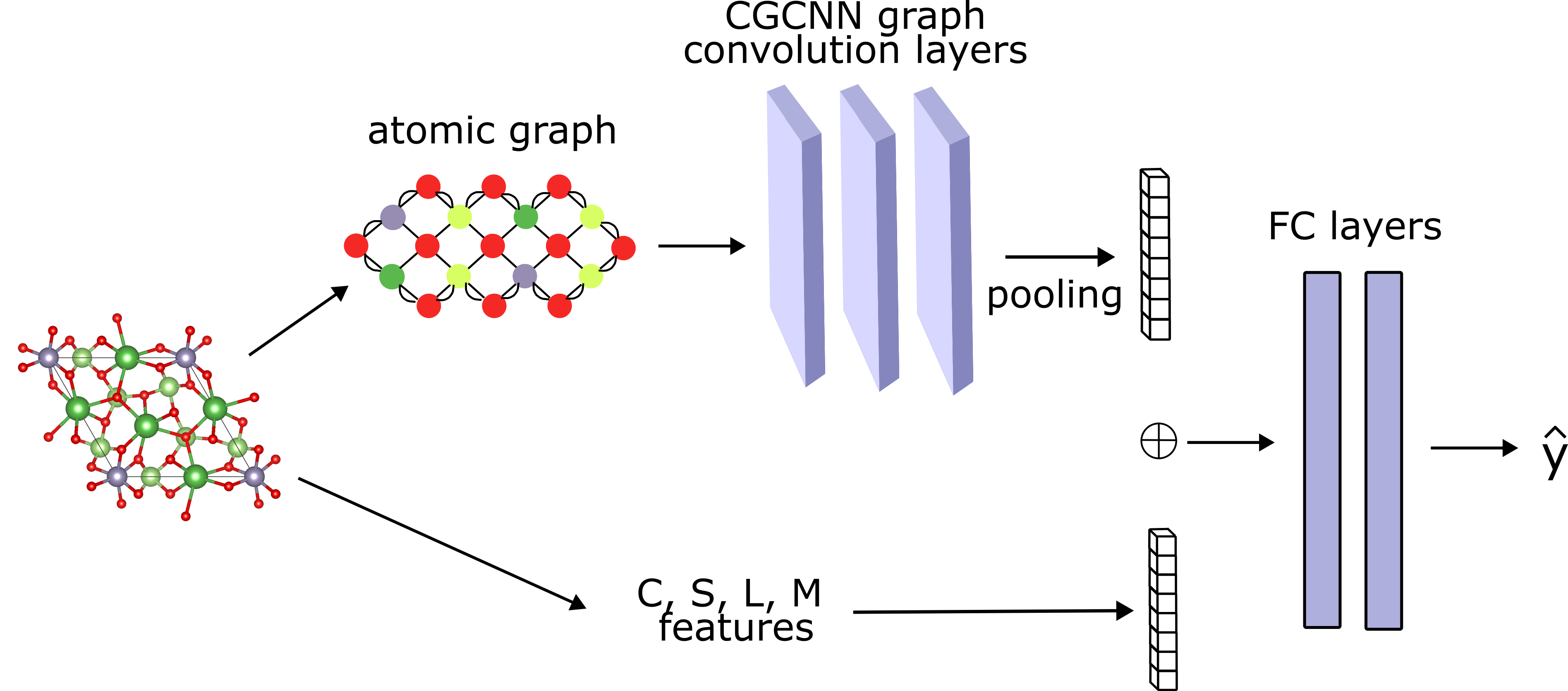}
  \caption{The schema of the CGCNNd model}
  \label{fgr:cgcnnd}
\end{figure}

As input features for structure-based GNN models, cgcnn features, extended cgcnn features (cgcnn+edc), and mat2vec features. 
In the work~\cite{Gong2023} it was shown that, due to the limited power of the considered GNNs in representation learning some properties of materials, which are commonly used as physical descriptors, the GNN models fail to learn. The authors performed a thorough analysis of the issue and advocated for injecting the descriptors that are difficult to learn in GNN models after graph convolution layers. These include mainly features related to periodicity and symmetry, such as unit cell parameters, space group, crystal system, etc. As these properties are important for the prediction of k-point distance, we use the modification suggested in~\cite{Gong2023} for our GNN models. Resulting models are named CGCNNd and ALIGNNd, respectively.  The schema CGCNNd model is shown in Figure~\ref{fgr:cgcnnd}. To inform GNN models about the metallic properties of the compounds, we also use our metallicity features in CGCNNd and ALIGNNd.

\subsection{Estimation of uncertainty}
Predicting k-point distance with machine-learning models inevitably introduces errors. These errors affect DFT calculations asymmetrically. Underestimation of the k-point distance and ensuing oversampling of the Brillouin zone, is safe, since the SCF calculation still lies in the converged region. Conversely, overestimation of k-distances iwould lead to undersampling of the Brillouin zone and unreliable results. To reduce the probability of overprediction we shift the prediction target from the median (50th percentile) to the lower conditional quantile of the k-point distance distribution.

Prediction uncertainty has two main sources:  (1) aleatoric uncertainty, which represents statistical noise in the data; (2) epistemic uncertainty, which arises due to uncertainty in the model with which we fit the data; this uncertainty arises due to the finite number of points in the training, or model misspecification. Since our objective is to estimate lower quantiles of the required k-point distance to avoid overestimation, we primarily focus on aleatoric uncertainty.

To estimate aleatoric uncertainty, we employ conformalised quantile regression ~\cite{romano2019conformalizedquantileregression}, which we found to be the most effective approach for our application. This method inherits the strongest features of the predecessor methods: quantile regression and conformal prediction. 

\textbf{Quantile regression} learns the conditional quantiles of a dependent variable as a linear function of the explanatory variables $Q_{\alpha}(x) = {y: F(y|x)\geq \alpha}$. To perform Quantile regression, one uses a special asymmetric, pinball loss function:
\begin{equation}
L_{\alpha}\bigl(y_i, \hat{y}_i(x_i)\bigr)
=
\begin{cases}
\alpha \, \bigl( y_i - \hat{y}_i(x_i) \bigr), & \text{if } y_i > \hat{y}_i(x_i), \\[6pt]
(1-\alpha) \, \bigl( \hat{y}_i(x_i) - y_i \bigr), & \text{if } y_i \le \hat{y}_i(x_i).
\end{cases}
\end{equation}
This way, one $\alpha$-quantile is learned. We use this loss to train ALIGNN and CGCNN models to predict conditional quantiles.

For Random Forest quantiles are estimated in another way~\cite{Meinshausen2006}. Conditional quantiles can be inferred with Quantile Regression Forests (QRF) is trained with a standard MSE loss, and is a generalisation of random forests, which retain for each leaf, the \textbf{empirical distribution} of training targets rather than only their mean. The empirical distribution is the weighted distribution of observed response variables, where the weights attached to observations are identical to the original random forest algorithm. And from the estimation of conditional distribution $\hat{F}(y|x=X)$, the $\alpha$-quantile $Q_{\alpha}(x)$:
\begin{equation}
    Q_{\alpha}(x)=\inf \{y:F(y|X=x)\geq \alpha\}
\end{equation}
We use the implementation of the Quantile Regression Forests from the sklearn-quantile package.

\textbf{Conformal prediction (CP)} is a distribution-free framework that provides valid prediction intervals with valid finite-sample marginal coverage under the assumption of exchangeability of samples. Unlike parametric methods, it does not rely on strong distributional assumptions and can be applied as a wrapper around any machine learning method. Given a regression model f(x), CP constructs a prediction interval:
\begin{equation}
C(x_{n+1})=[f(x_{n+1})-\beta_{\alpha}, f(x_{n+1})+\beta_{\alpha}]
\end{equation}
where:
\begin{equation}
\beta_{\alpha} = Q_{(1-\alpha)(1+1/|C|)}(\{\beta_i\} for i \in C)
\end{equation}
is an empirical quantile of residuals $\beta_i=|y_i-f(x_i)|$ on the calibration set C.

\textbf{Conformalised Quantile Regression (CQR)} preserves the strong features of quantile regression and conformal predictions, and provides theoretical guarantees of distribution-free finite-sample coverage with asymmetric, heteroscedastic intervals conditioned on data points.  CQR begins with a model that predicts lower and upper conditional quantiles $Q_{\alpha\_lo}(x)$ and $Q_{\ alpha\_hi}(x)$. Then, conformity scores are calculated on calibration dataset C: $\beta_i = \max\{Q_{\alpha\_lo}(X_i)-Y_i, Y_i-Q_{\alpha\_hi}(X_i)\}$. Then the empirical quantile $\beta_\alpha$ is calculated for the distribution of conformity scores as in CP method. And the final confidence intervals are determined as:
\begin{equation}
C(X_{n+1})=[Q_{\alpha\_lo}(X_{n+1})-\beta_\alpha, Q_{\alpha\_hi}(X_{n+1})+\beta_\alpha]
\end{equation}

\section{Results}
\subsection{Overall comparison of performance of models}
First, we train the models in combination with different sets of features, and the ways to construct atomic graph to predict the mean value of k-points distance. For all models with split data into training: validation: testing datasets in a 80:10:10 proportion. We perform random search over hyperparameter space on validation dataset for all models. Then, for the best set of hyperparameters, we perform manual hyperparameter search in the vicinity of it. For the final hyperparameter sets, all models are trained on 3-5 different data splits determined for randomly chosen random seeds, and their performance is averaged. Confidence intervals for performance are estimated under the assumption of a Gaussian distribution of errors in performance metrics. 

We evaluate model performance using mean absolute error (MAE), mean absolute percentage error (MAPE), mean squared error (MSE), the coefficient of determination ($R^2$), the Spearman rank correlation coefficient, and the Kendall rank correlation coefficient. The inclusion of rank-based metrics (Spearman and Kendall) is motivated by the need to assess how well the models preserve the correct pairwise ordering of k-point densities. This allows us to quantify whether the ML models introduce additional randomness in the relative ordering of samples, even when absolute errors remain small.

\begin{figure}[h]
\centering
  \includegraphics[width=0.6\columnwidth]{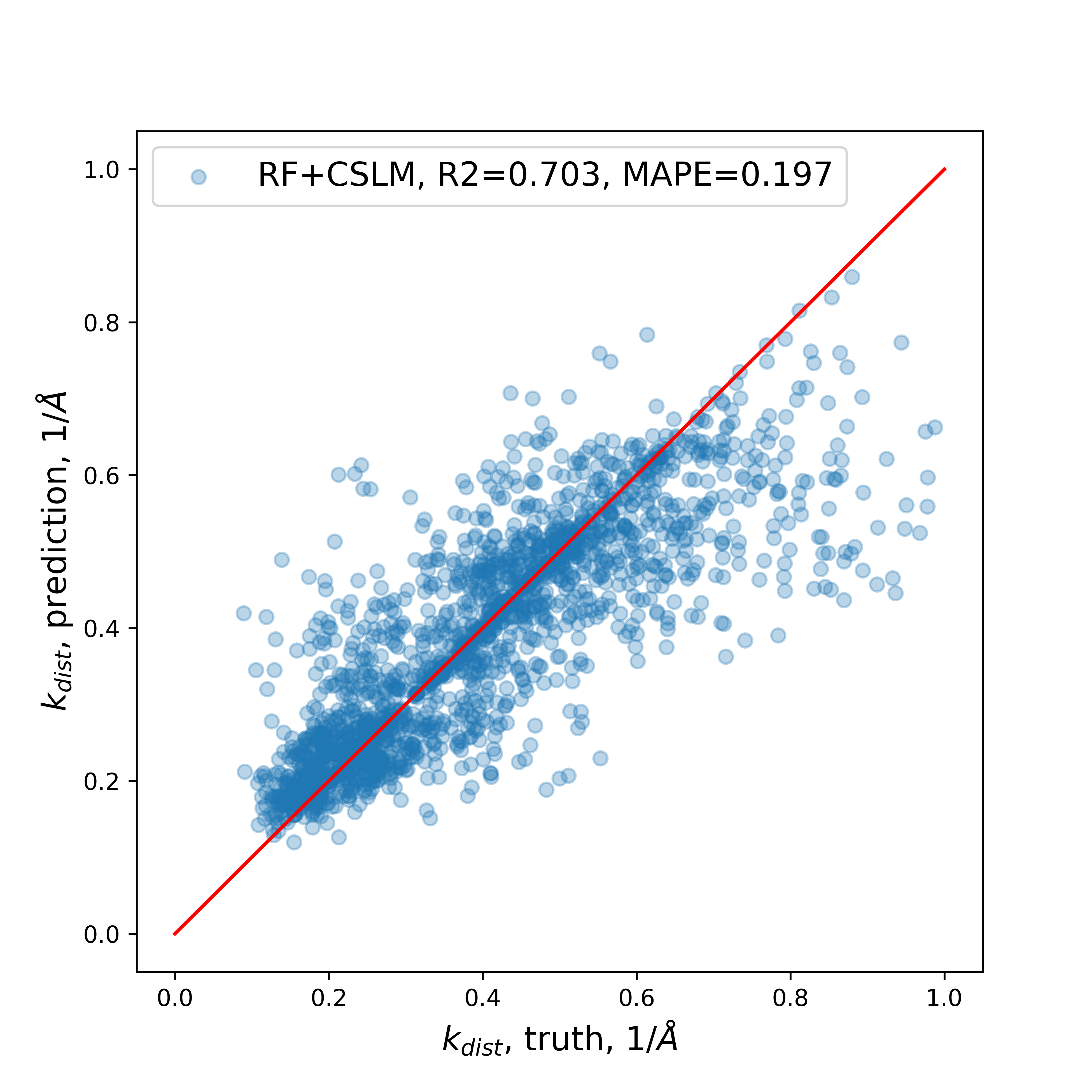}
  \caption{Truth vs prediction for the best ensemble model RF+CSLM on the test set}
  \label{fgr: RF-best}
\end{figure}

Table~\ref{tbl:ensemble-models} shows the performance of the ensemble models, RF and GB,  trained with different feature sets. In the table, feature sets are C=composition features (size=146), CS=composition+structure+SOAP (size=404), CSL= composition+structure+SOAP+lattice (size=419), CSLM= composition+structure+SOAP+lattice+metallicity (size=483), JarvisCFID (size=1557). One can see that the performance of Random Forest in general is better than of GB on this task, and the best feature set is CSLM independently of the model. The best model, Random Forest + CSLM, achieves an $R^2$ score of ~ 0.703, MAE = 0.067, MAPE = 0.197, Spearman correlation = 0.86. Figure~\ref{fgr: RF-best} shows the scatter plot for predictions made by this model.

\begin{table*}
\caption{Performance of different ensemble models}
\label{tbl:ensemble-models}
\setlength{\tabcolsep}{3pt}
\begin{ruledtabular}
\scriptsize
\begin{tabular}{lllllllll}
\# & Model & Features & MAE & MAPE & MSE & $R^2$ & Spearman & Kendall \\
\hline
1 & RF & C & $0.075 \pm 0.0004$ & $0.221 \pm 0.003$ & $0.0109 \pm 0.0002$ & $0.641 \pm 0.005$ & $0.821 \pm 0.004$ & $0.629 \pm 0.002$ \\
2 & RF & CS & $0.0695 \pm 0.0006$ & $0.206 \pm 0.005$ & $0.0095 \pm 0.0001$ & $0.686 \pm 0.008$ & $0.850 \pm 0.004$ & $0.661 \pm 0.005$ \\
3 & RF & CSL & $0.067 \pm 0.002$ & $0.197 \pm 0.006$ & $0.0091 \pm 0.0003$ & $0.702 \pm 0.009$ & $0.861 \pm 0.006$ & $0.674 \pm 0.008$ \\
4 & RF & CSLM & $0.067 \pm 0.001$ & $0.197 \pm 0.006$ & $0.0090 \pm 0.0003$ & $0.703 \pm 0.008$ & $0.861 \pm 0.006$ & $0.676 \pm 0.006$ \\
5 & RF & JarvisCFID & $0.078 \pm 0.0007$ & $0.234 \pm 0.005$ & $0.0113 \pm 0.0002$ & $0.628 \pm 0.003$ & $0.815 \pm 0.004$ & $0.618 \pm 0.004$ \\
6 & GB & C & $0.080 \pm 0.002$ & $0.240 \pm 0.005$ & $0.0118 \pm 0.0004$ & $0.616 \pm 0.007$ & $0.804 \pm 0.004$ & $0.605 \pm 0.004$ \\
7 & GB & CS & $0.0741 \pm 0.0008$ & $0.219 \pm 0.004$ & $0.0102 \pm 0.0003$ & $0.659 \pm 0.009$ & $0.833 \pm 0.005$ & $0.636 \pm 0.005$ \\
8 & GB & CSL & $0.0731 \pm 0.0008$ & $0.214 \pm 0.002$ & $0.0100 \pm 0.0003$ & $0.668 \pm 0.008$ & $0.839 \pm 0.005$ & $0.642 \pm 0.006$ \\
9 & GB & CSLM & $0.074 \pm 0.002$ & $0.218 \pm 0.005$ & $0.0102 \pm 0.0005$ & $0.660 \pm 0.010$ & $0.833 \pm 0.008$ & $0.640 \pm 0.008$ \\
10 & GB & JarvisCFID & $0.0803 \pm 0.0008$ & $0.236 \pm 0.005$ & $0.0119 \pm 0.0004$ & $0.610 \pm 0.010$ & $0.807 \pm 0.008$ & $0.606 \pm 0.006$ \\
\end{tabular}
\end{ruledtabular}
\end{table*}

Comparing the performance of the ensemble models with different feature sets, we can see that each subset of features contributes to the final result. To quantify the contribution of each block of features to the performance of the final model, we make predictions on modified representations of samples from the test set. We shuffle features in each block of features and see how much the performance of the prediction drops. Results are shown in Table~\ref{tbl:feature-importance}. It can be seen that each set of features contributes to the final prediction, with composition features having the largest contribution, followed by the metallicity features and lattice features. From the viewpoint of general physical/chemistry knowledge, this rating seems to be reasonable and supports the considerations we used when constructing the list of features

\begin{table}
\caption{Performance of RF+CSLM on test set described by modified features.
Each feature modification represents feature shuffling in one of the feature blocks.}
\label{tbl:feature-importance}
\setlength{\tabcolsep}{3pt}
\begin{ruledtabular}
\scriptsize
\begin{tabular}{llllllll}
\# & \makecell{Shuffled\\block} & MAE & MAPE & MSE & $R^2$ & $\Delta$MAE & $\Delta R^2$ \\
\hline
1 & -- & 0.067 & 0.197 & 0.009 & 0.703 & 0     & 0     \\
2 & composition  & 0.102 & 0.347 & 0.017 & 0.455 & 0.035 & 0.248 \\
3 & structure    & 0.068 & 0.192 & 0.009 & 0.693 & 0.001 & 0.010 \\
4 & SOAP         & 0.074 & 0.211 & 0.011 & 0.644 & 0.007 & 0.059 \\
5 & lattice      & 0.074 & 0.204 & 0.011 & 0.626 & 0.007 & 0.077 \\
6 & metallicity  & 0.078 & 0.248 & 0.011 & 0.612 & 0.011 & 0.091 \\
\end{tabular}
\end{ruledtabular}
\end{table}

Next, we turn to the training of the CrabNet model, which is based on transformer architecture and takes composition information as the only input.  CrabNet was trained with mat2vec and extended cgcnn features (cgcnn+edc). Notably, the performance of this composition-only model is significantly worse than for models with some encoding of structural information. For k-point prediction models, one would expect this kind of behaviour for models that capture the essence of the task.

\begin{table*}
    \caption{Performance of CrabNet models}
    \label{tbl:ensemble-models}
    \setlength{\tabcolsep}{3pt}
    \begin{ruledtabular}
    \scriptsize
  \begin{tabular*}{\textwidth}{@{\extracolsep{\fill}}lllllllll}
    \# & Model & Features & MAE & MAPE & MSE  & $R^2$ & Spearman & Kendall\\
    \hline
    1 & CrabNet & mat2vec & $0.084 \pm 0.001$ & $0.248 \pm 0.007$ & $0.0134 \pm 0.0003$  & $0.56 \pm 0.01$ & $0.772 \pm 0.005$ & $0.575 \pm 0.005$\\
    2 & CrabNet & cgcnn+edc & $0.084 \pm 0.002$ & $0.245 \pm 0.004$ & $0.0138 \pm 0.0006$  & $0.55 \pm 0.02$ & $0.77 \pm 0.01$ & $0.58 \pm 0.01$\\
  \end{tabular*}
  \end{ruledtabular}
\end{table*}

Table~\ref{tbl:gnn-models} shows the performance of GNN models. CGCNN model was trained with cgcnn features and two ways to construct atomic graph, standard radius graph (rg+cgcnn) and crystalline graph, which has edges only between atoms sharing Voronoi surfaces (crg+cgcnn). The standard radius graph performs better. Because of it we used standard radius graph in all other cases. Next, we try different atomic input features for CGCNN, the list includes cgcnn (rg+cgcnn), mat2vec (rg+mat2vec), cgcnn+edc (rg+cgcnn+edc). From all these combinations standard cgcnn features give best performance. So, for this combination we try to inject additional features after GNN layers. The schema of the extended CGCNN model in shown in Figure~\ref{fgr:cgcnnd}. The performance of extended CGCNNd model is slightly improved compared to vanilla model, both in case of CSL combination of features (CSL = composition + structure (without SOAP, just matminer) + lattice) and metallicity features (m).
The ALIGNN model was trained with radius graph only (for construction of both atomic and linear graph) with cgcnn and cgcnn+edc features, with the latter showing slightly better performance. So, we choose this combination for the extended ALIGNNd model taking in additionally CSL or metallicity features. The performance of ALIGNNd+cgcnn+edc+CSL model is comparable to our best Random Forest model.

\begin{table*}
    \caption{Performance of the GNN models}
      \label{tbl:gnn-models}
    \setlength{\tabcolsep}{3pt}
    \begin{ruledtabular}
    \scriptsize
  \begin{tabular*}{\textwidth}{@{\extracolsep{\fill}}lllllllll}
    \# & Model & Features & MAE & MAPE & MSE  & $R^2$ & Spearman & Kendall\\
    \hline
    1 & CGCNN  & rg+cgcnn            & $0.074 \pm 0.002$  & $0.207 \pm 0.004$ & $0.0112 \pm 0.0006$ & $0.630 \pm 0.010$ & $0.827 \pm 0.006$ & $0.634 \pm 0.008$\\
    2 & CGCNN  & crg+cgcnn           & $0.077 \pm 0.001$  & $0.223 \pm 0.006$ & $0.0115 \pm 0.0003$ & $0.630 \pm 0.010$ & $0.819 \pm 0.006$ & $0.626 \pm 0.005$\\
    3 & CGCNN  & rg+mat2vec          & $0.0748 \pm 0.0008$ & $0.208 \pm 0.005$ & $0.0116 \pm 0.0002$ & $0.620 \pm 0.005$ & $0.822 \pm 0.003$ & $0.629 \pm 0.003$\\
    4 & CGCNN  & rg+cgcnn+edc        & $0.075 \pm 0.002$  & $0.215 \pm 0.004$ & $0.012 \pm 0.001$  & $0.630 \pm 0.020$ & $0.819 \pm 0.007$ & $0.629 \pm 0.008$\\
    5 & ALIGNN & rg+cgcnn            & $0.0700 \pm 0.0005$ & $0.200 \pm 0.001$ & $0.0105 \pm 0.0002$ & $0.644 \pm 0.008$ & $0.837 \pm 0.003$ & $0.650 \pm 0.003$\\
    6 & ALIGNN & rg+cgcnn+edc        & $0.0717 \pm 0.0007$ & $0.194 \pm 0.003$ & $0.0106 \pm 0.0002$ & $0.654 \pm 0.008$ & $0.845 \pm 0.004$ & $0.656 \pm 0.004$\\
    7 & CGCNNd & rg+cgcnn+CSL        & $0.0737 \pm 0.0007$ & $0.206 \pm 0.005$ & $0.0111 \pm 0.0004$ & $0.647 \pm 0.005$ & $0.841 \pm 0.001$ & $0.649 \pm 0.001$\\
    8 & CGCNNd & rg+cgcnn+M          & $0.0740 \pm 0.0002$ & $0.209 \pm 0.001$ & $0.0117 \pm 0.0008$ & $0.630 \pm 0.020$ & $0.839 \pm 0.004$ & $0.648 \pm 0.003$\\
    9  & ALIGNNd & rg+cgcnn+edc+CSL   & $0.069 \pm 0.002$  & $0.186 \pm 0.004$ & $0.0099 \pm 0.0003$ & $0.680 \pm 0.010$ & $0.856 \pm 0.005$ & $0.669 \pm 0.008$\\
    10 & ALIGNNd & rg+cgcnn+edc+M     & $0.070 \pm 0.001$  & $0.190 \pm 0.004$ & $0.0104 \pm 0.0002$ & $0.662 \pm 0.008$ & $0.849 \pm 0.003$ & $0.661 \pm 0.004$\\
  \end{tabular*}
  \end{ruledtabular}
\end{table*}

\begin{figure}[h]
\centering
  \includegraphics[width=0.6\columnwidth]{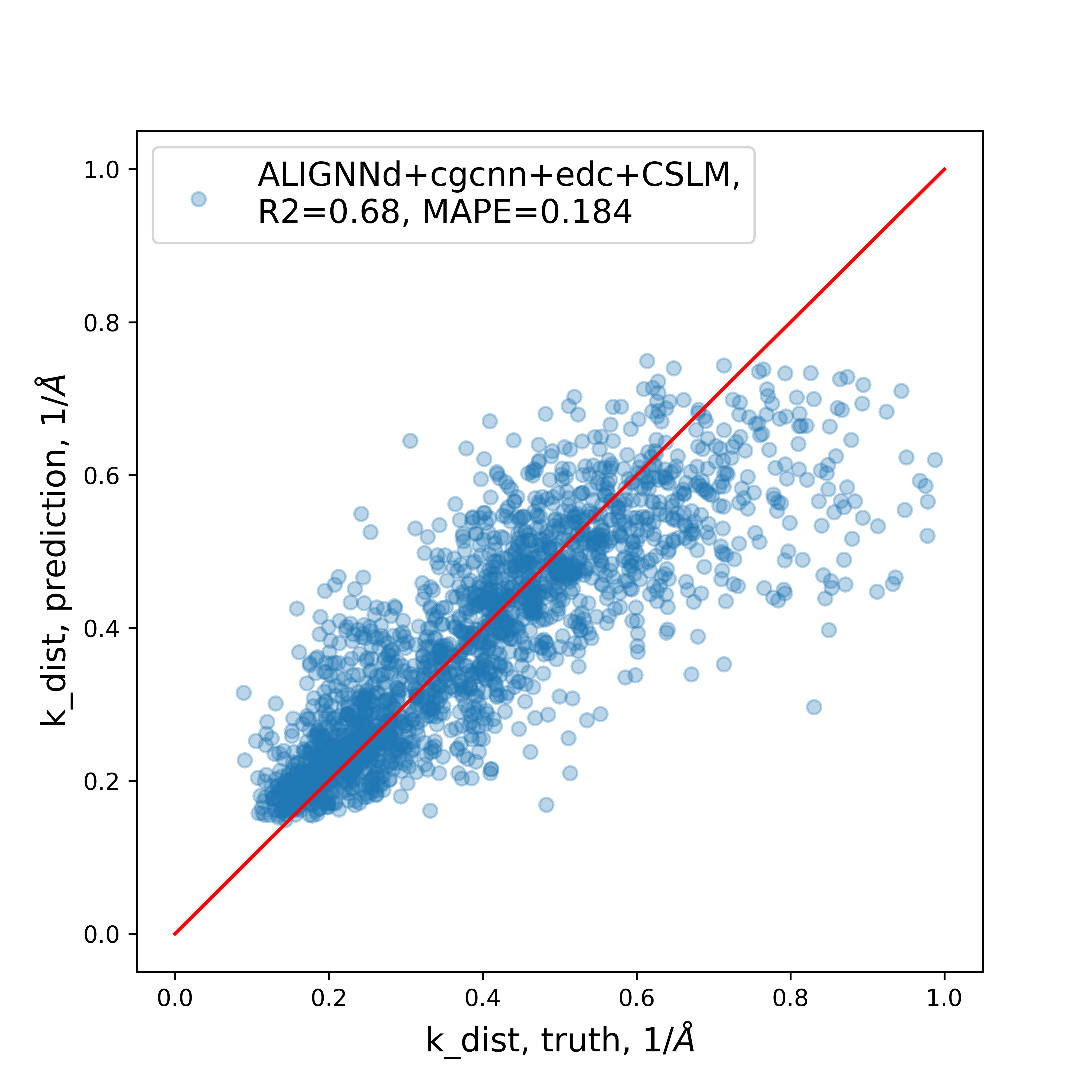}
  \caption{Truth vs prediction ALIGNNd+cgcnn+edc+CSLM on the test set.}
  \label{fgr:alignnd+CSLM}
\end{figure}

Across all models tested, the ensemble methods (Random Forest, Gradient Boosting) outperform composition-only neural models (CrabNet) while performing comparably to advanced graph neural networks (ALIGNN). The best-performing model overall is RF+CSLM, which reaches an R² of ~0.7, slightly surpassing ALIGNNd+cgcnn+edc+CSL model. 

\subsection{Interpretation of k-points prediction results}
To analyse the factors affecting the decision about k-points distance value, we train a surrogate model~\cite{molnar2025}, which is a shallow decision tree on the predictions of the RF+CSLM model. As feature vectors are very high-dimensional, we do this in two stages. First, we train RF+CSLM on different data splits, and for each of them, fit a surrogate model. These decision trees have a lot in common, but also some features that are different between different data splits. We collect all features that appear in those surrogate decision trees and reduce the features to only those that appear in the surrogate model in composition, structure, and lattice block. We then perform PCA reduction to 5 features for the SOAP and CGCNN block. The new features have a size 20 instead of 483. Next, we refit the surrogate decision trees on these reduced feature sets to obtain simple sets of decision rules for k-point distance prediction. The $R^2$-score for the final surrogate model with depth 4 and minimum 40 samples per leaf is 0.706 (we compare predictions of the surrogate model with predictions of the fitted RF+CSLM model here, not the true values).

\begin{figure*}
 \centering
 \includegraphics[height=6cm]{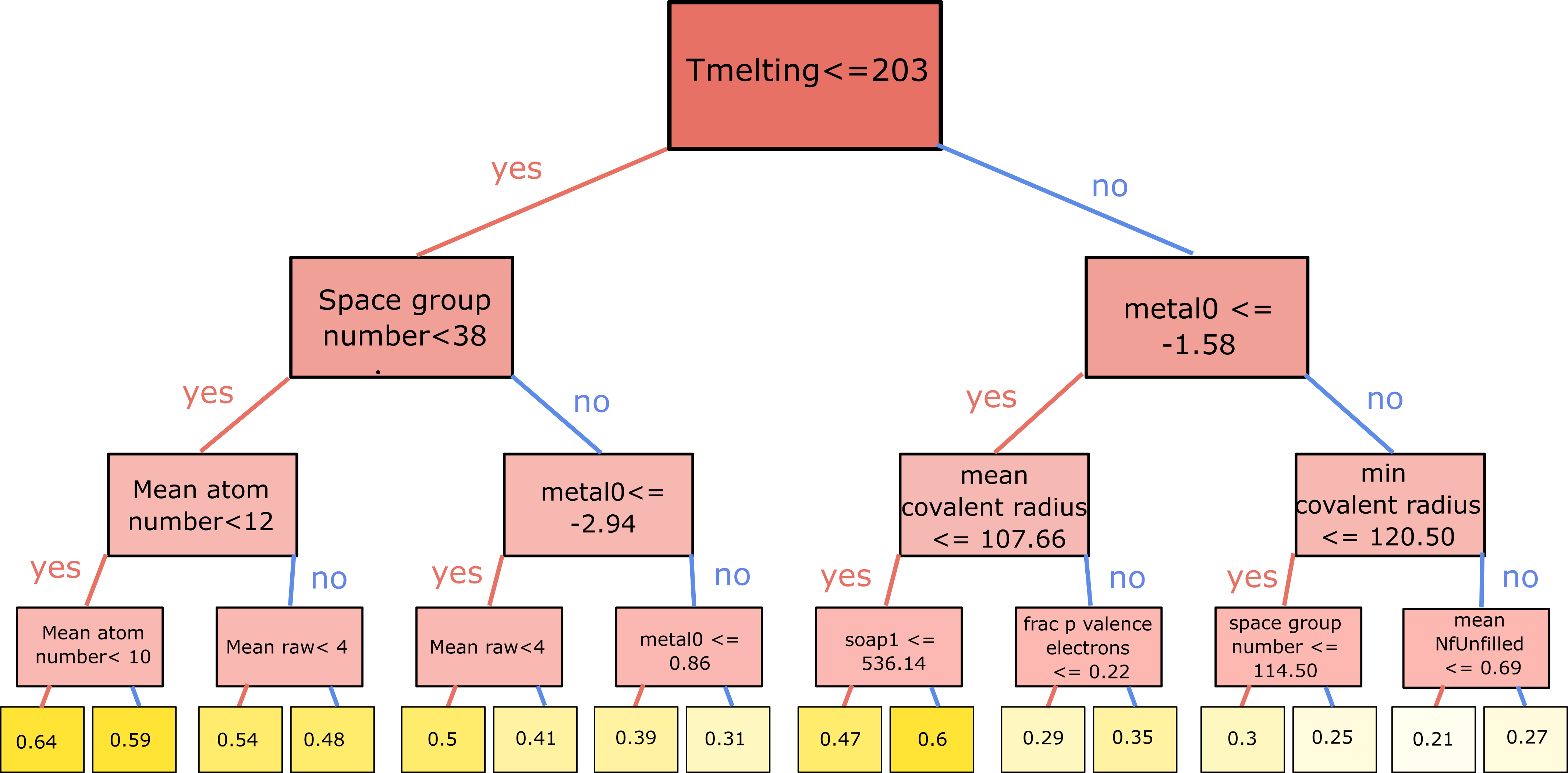}
 \caption{Surrogate model for RF-CSLM predictions. Yellow rectangles are leaves with final values of k-point distance.}
 \label{fgr:RF-surrogate}
\end{figure*}

An example of the final decision tree is shown in Figure~\ref{fgr:RF-surrogate}. One can see that the root is a melting temperature. Melting temperature is not a true melting temperature, but a feature which is calculated as a weighted mean of melting temperatures of atomic compounds composed of atoms in the composition formula, and the weights are calculated from atomic fractions in the composition. And the splitting value is 203 K. It is an extremely low melting temperature; the elements with melting temperatures below this threshold are H, N, O, F, Cl, Ne, Ar, Kr, Xe, Rn. So, the compounds with melting temperature below 203 are those rich in these elements, and surprisingly, there is approximately 50\% of such compounds in the dataset. Splitting at the melting temperature is a very stable feature appearing in all decision trees that we built for different splits of data.

Decision subtrees for compounds with melting temperatures below 203 K, and above this value, are very different. Following the ‘yes’ branch, we arrive at the condition ‘space group number < 38’. This is also a very stable feature that persists for all data splits. For a low-symmetry compound, the next feature to split upon is the mean atom number. For high symmetry compounds, the split on the third level of the tree is on the first PCA component (most informative) of the metallicity vector. Meatl0 feature carry information about the metallicity of the compound. However, the correspondence of the metallicity property of the compound is not fully described by this feature alone, and the meaning of the numerical value can’t be interpreted easily. The final level of decision making is composed of mean atomic number, mean atomic raw, and metal0.

Following the ‘no’ branch, we arrive and the condition metal0 $\leq$ -1.58. Next level of the decision tree uses “mean covalent radius” or “min covalent radius” depending on the metallicity property, showing the importance of the covalent radius in decision making about k-points distance. The final level of decision-making is composed of structural and electronic features. 

Analysis of decision making by the RF-CSLM model shows that the k-points distance is determined by diverse behaviour, and apart from composition features, global symmetry descriptors, metallicity descriptors, and electronic configuration are important too. This conclusion additionally explains why CrabNet is so inefficient for this task, and why GNN models perform worse than RF+CSLM.

\subsection{Uncertainty estimation results}
After identifying the best models, we perform uncertainty estimation for these models with conformalised quantile regression. Performing uncertainty estimation allows us to shift the prediction from the mean value to the $\alpha$-quantile, ensuring that at least $\alpha \cdot 100\%$ of cases of predicted k-points distances correspond to dense enough meshes to ensure convergence. To evaluate the models making shifted predictions, we establish a baseline as a constant value of k-points distance, chosen in a way that it is the largest possible k-points distance for which $\alpha \cdot 100\%$ of compounds are converged. This value is expected to be different depending on the chosen confidence level. We would like to stress that usually the training set is not available or very small, so the constant value of k-points for high-throughput calculations is chosen in a less rigorous way. So, the suggested baseline is strong. The mean error, showing the deviation of the used value from the true one, represents the measure of the accuracy of the models and translates into computational time in practical settings. Smaller errors correspond to more accurate k-point distances prediction and more computation time saved.

\begin{figure}[h]
\centering
  \includegraphics[width=0.6\columnwidth]{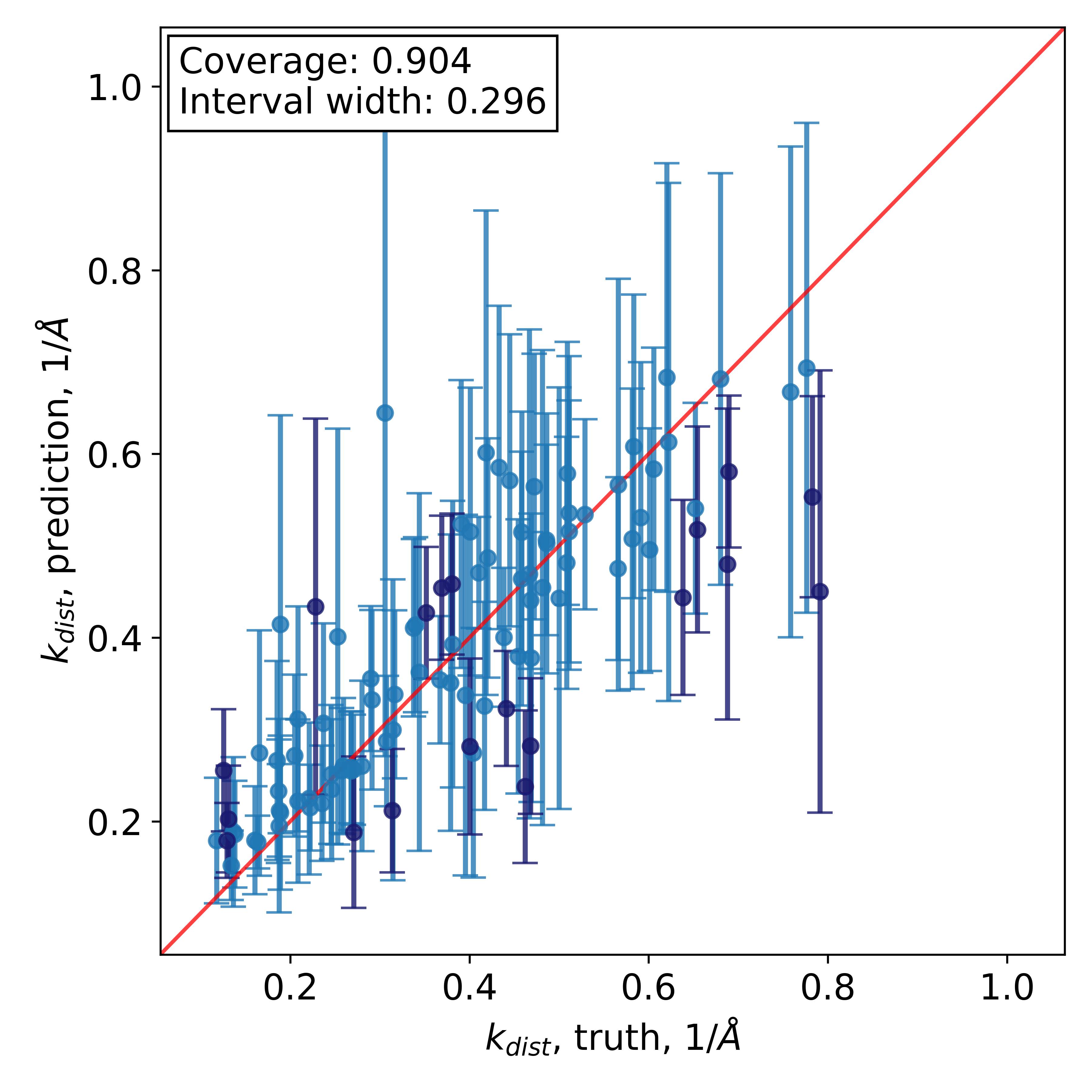}
  \caption{CQR for ALIGNNd+cgcnn+edc+CSLM, target coverage 0.9.}
  \label{fgr:alignnd-cqr-90}
\end{figure}

\begin{figure}[h]
\centering
  \includegraphics[width=0.6\columnwidth]{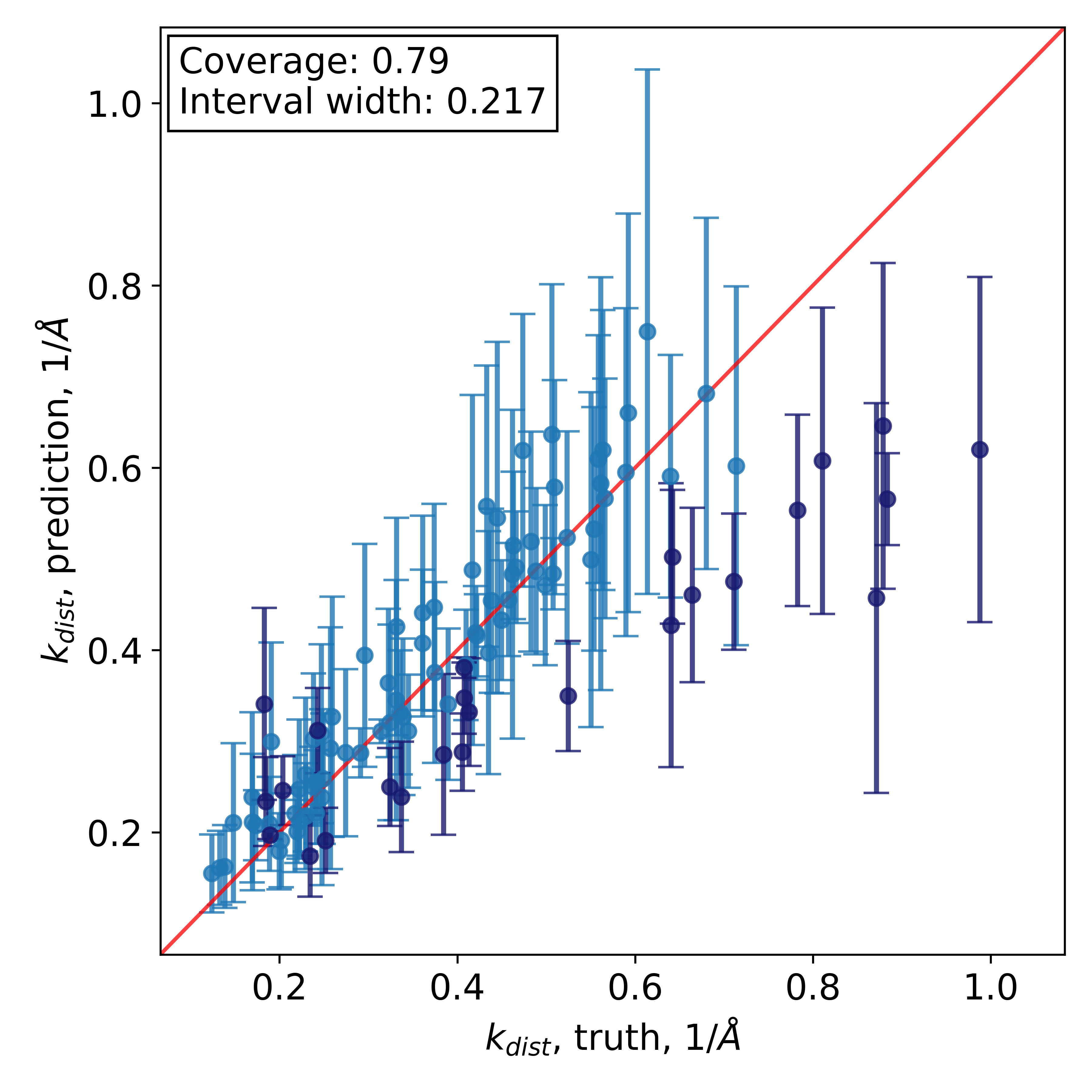}
  \caption{CQR for ALIGNNd+cgcnn+edc+CSLM, target coverage 0.8.}
  \label{fgr:alignnd-cqr-80}
\end{figure}

\begin{figure}[h]
\centering
  \includegraphics[width=0.6\columnwidth]{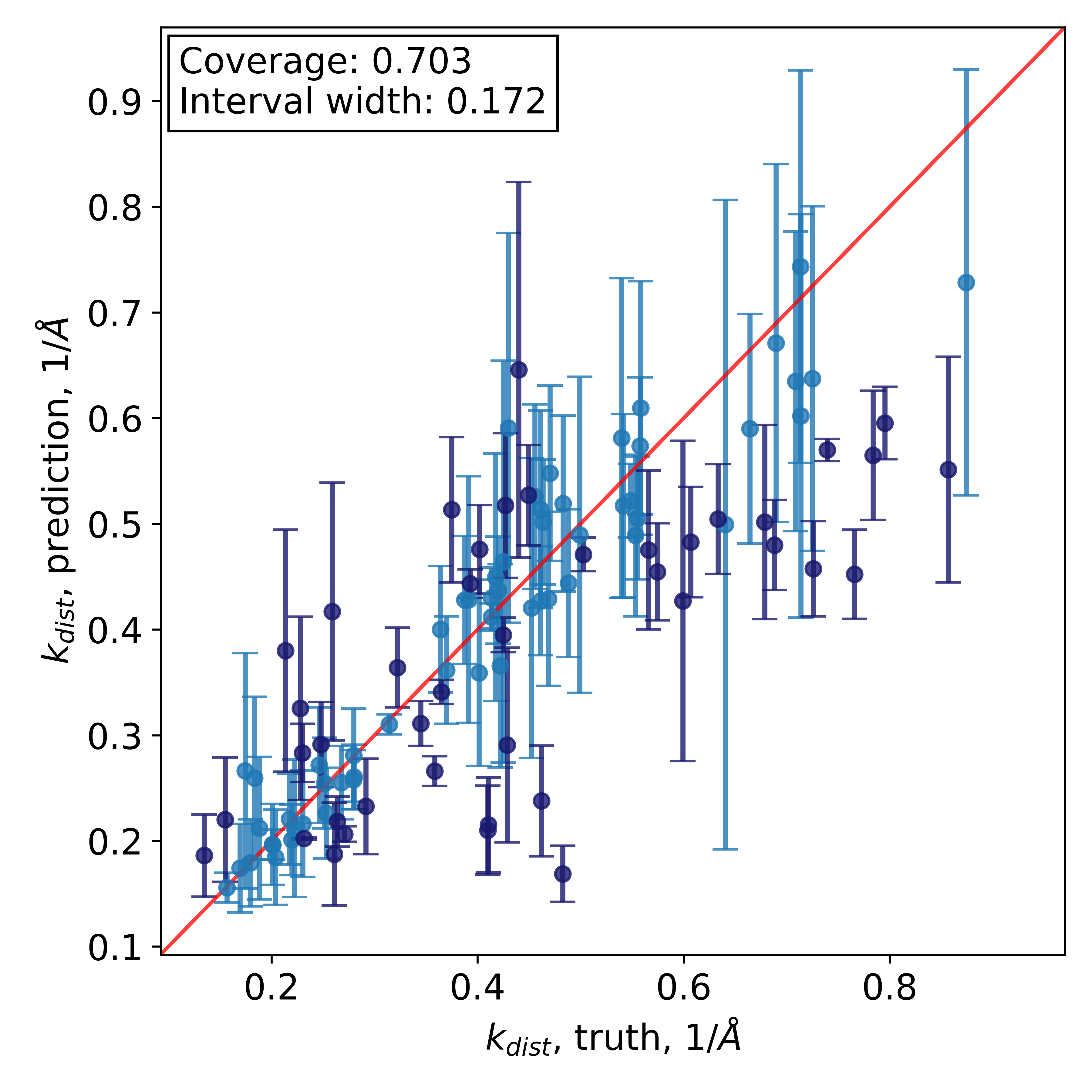}
  \caption{CQR for ALIGNNd+cgcnn+edc+CSLM, target coverage 0.7.}
  \label{fgr:alignnd-cqr-70}
\end{figure}

The results of CQR for ALIGNNd+cgcnn+edc+CSLM are shown in Figures~\ref{fgr:alignnd-cqr-90},~\ref{fgr:alignnd-cqr-80},~\ref{fgr:alignnd-cqr-70} for CQR coverage 0.9, 0.8, and 0.7. Light blue points and confidence intervals show samples for which the true value is inside that confidence interval; dark blue points and confidence intervals show samples for which the true value lies outside of the confidence interval. It can be seen that the CQR empirical coverage corresponds well to the target coverage. The mean size of the confidence intervals decreases with a reduction in coverage, as expected.

Table~\ref{tbl:cqr-performance} shows the results of conformalised quantile regression for all best models. In all cases, the empirical coverage is consistent with the target coverage, and the confidence intervals have a reasonable size. This means that our models guarantee that a target fraction of compounds is calculated accurately, conditioned on training set distribution. ALIGNNd+cgcnn+edc+CSLM provides slightly better coverage, narrower confidence intervals, and correspondingly smaller errors than RF+CSLM. All ML models at all confidence levels provide smaller overall errors than the baseline model. So, using ML models for the prediction of k-point distances is practically beneficial. 

\begin{table*}
  \caption{Performance of CQR models using different target coverages and target quantiles. 
  Empirical coverage and MAE values are computed on the test set.}
  \label{tbl:cqr-performance}
  \setlength{\tabcolsep}{3pt}
    \begin{ruledtabular}
    \scriptsize
  \begin{tabular*}{\textwidth}{@{\extracolsep{\fill}}lllllllllll}
    \hline
    \# & Model & Features & \makecell{Target\\coverage} & \makecell{Target\\quantile} & \makecell{CQR\\interval\\coverage} & \makecell{CQR\\mean\\width} & \makecell{Empirical\\coverage} & MAE & \makecell{k-distance\\constant\\baseline} & \makecell{MAE\\baseline} \\
    \hline
    1 & RF      & CSLM                      & 0.95 & 0.90 & 0.895 & 0.313 & 0.958 & 0.152  & 0.158 & 0.237 \\
    2 & RF      & CSLM                      & 0.90 & 0.80 & 0.820 & 0.268 & 0.917 & 0.122  & 0.184 & 0.215 \\
    3 & RF      & CSLM                      & 0.85 & 0.70 & 0.720 & 0.180 & 0.849 & 0.104 & 0.205 & 0.200 \\
    4 & ALIGNNd & rg+cgcnn+edc+CSL          & 0.95 & 0.90 & 0.904 & 0.254 & 0.966 & 0.143 & 0.158 & 0.238 \\
    5 & ALIGNNd & rg+cgcnn+edc+CSL          & 0.90 & 0.80 & 0.790 & 0.217 & 0.908 & 0.114 & 0.184 & 0.215 \\
    6 & ALIGNNd & rg+cgcnn+edc+CSL          & 0.85 & 0.70 & 0.703 & 0.172 & 0.840 & 0.095 & 0.205 & 0.200 \\
  \end{tabular*}
  \end{ruledtabular}
\end{table*}

\begin{figure}[h]
\centering
  \includegraphics[width=0.6\columnwidth]{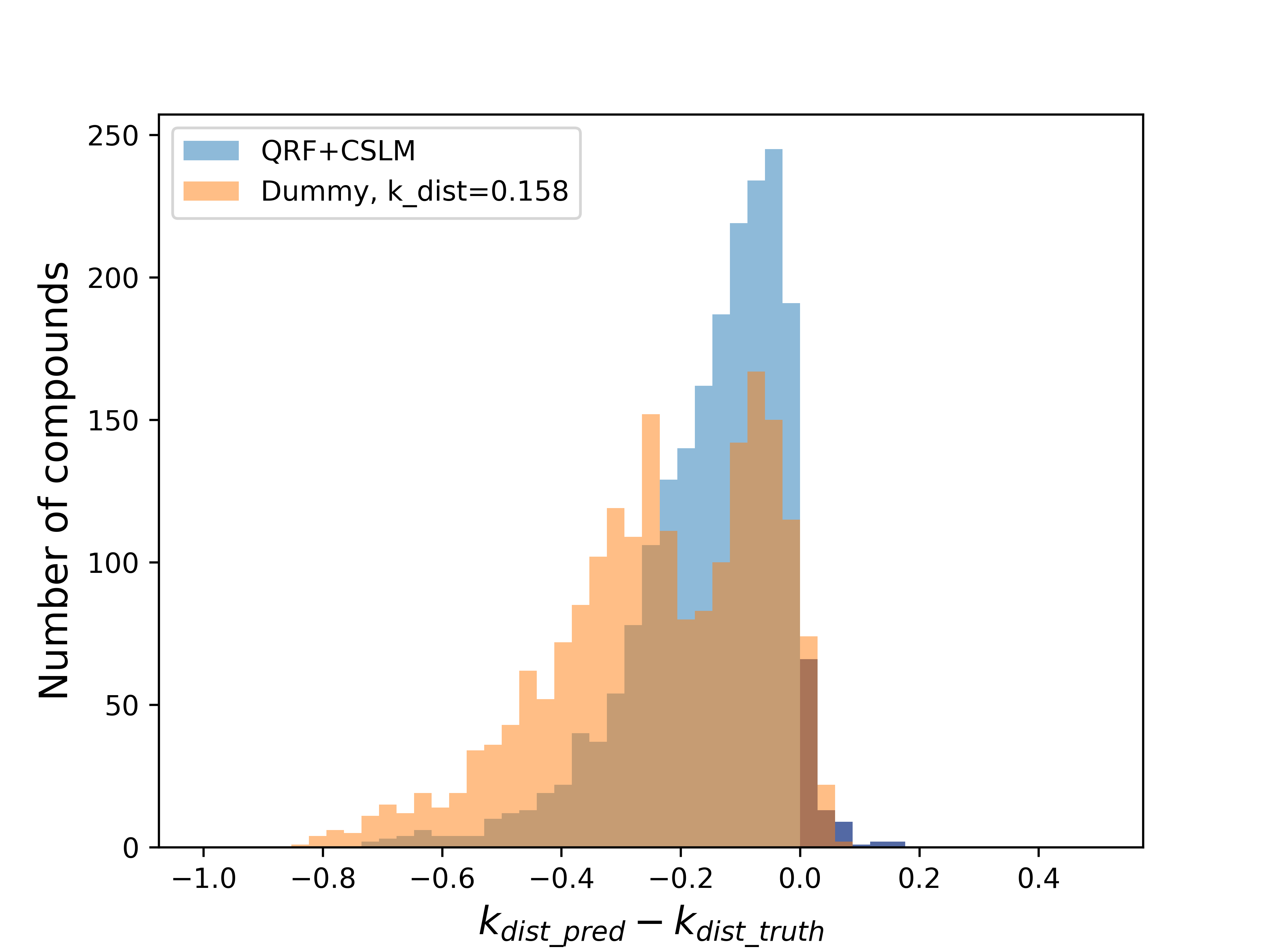}
  \caption{Distribution of errors for the shifted RF+CSLM model and the baseline model. Target coverage is 95\% of k-point distances. Dark blue bins show the samples for the predicted k-distance is larger than the true one.}
  \label{fgr:rf+baseline}
\end{figure}

\begin{figure}[h]
\centering
  \includegraphics[width=0.6\columnwidth]{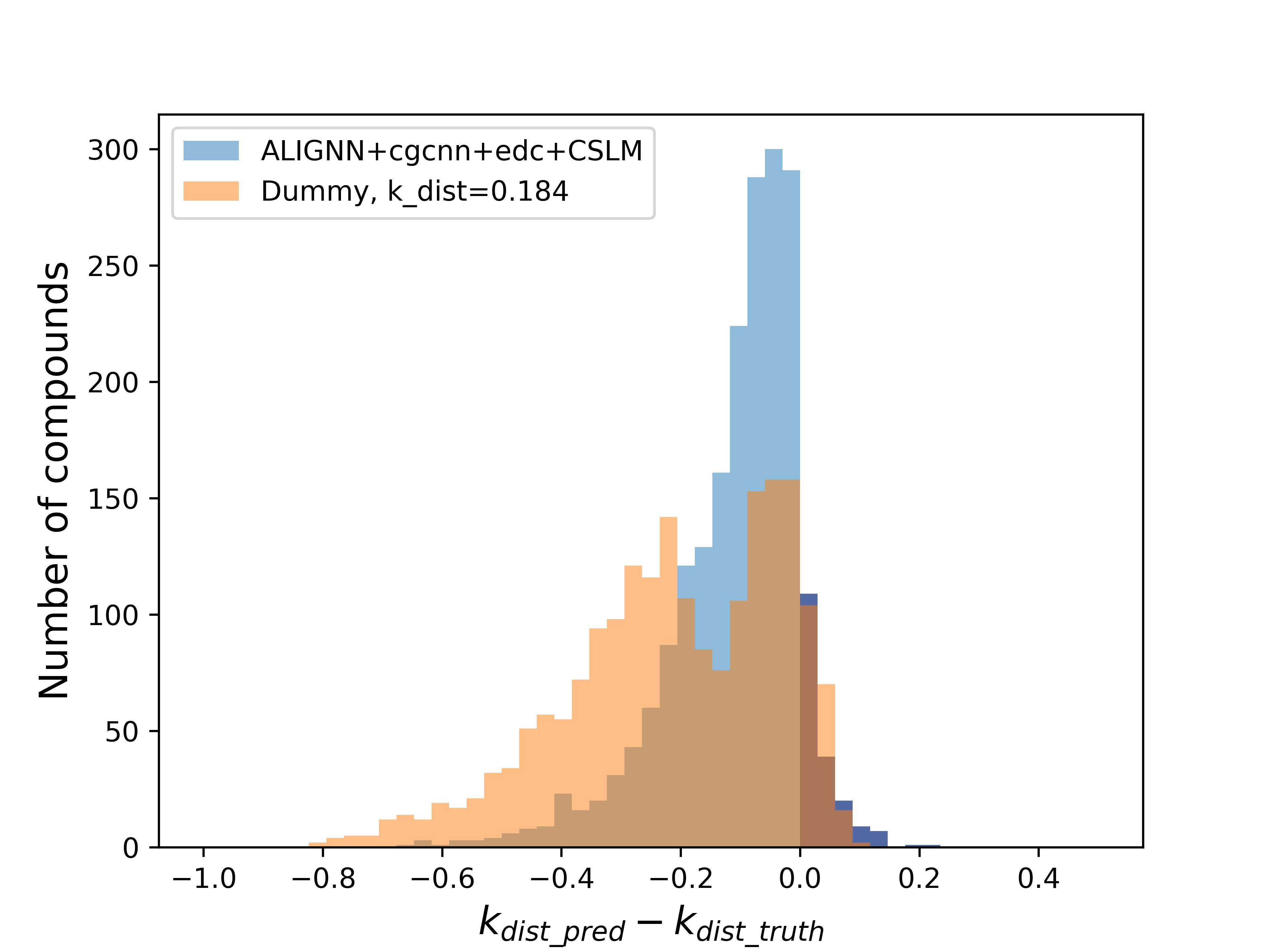}
  \caption{Distribution of errors for the shifted ALIGNN+cgcnn+edc+CSLM model and the baseline model. Target coverage is 90\% of k-point distances. Dark blue bins show the samples for the predicted k-distance is larger than the true one.}
  \label{fgr:alignnd+baseline}
\end{figure}

Figures~\ref{fgr:rf+baseline},~\ref{fgr:alignnd+baseline} show the distribution of errors for RF+CSLM (5th percentile), \\
ALIGNN+cgcnn+edc+CSLM (10th percentile), and the baseline model for different quantiles. Dark blue colour shows the samples for which the predicted k-distance is larger than the true one. These figures show that the distribution of errors for ML models is narrower than for the baseline model, which results in smaller mean absolute errors.


Translation of the gain in k-distance prediction error into gains in computation time in general will depend on a lot of details, such as parallelisation strategy, hardware setup, materials calculated, etc. However, to demonstrate that the gain is present, we calculate the gain in terms of wall time using data that we obtained in the process of generating our dataset. Results are shown in Table~\ref{tbl:walltime-analysis}. For our dataset, the mean wall time per sample is 352s, which is calculated taking the wall time corresponding to true converged k-distances. Both ML models adjusted to predict a quantile suitable for the required confidence, and the baseline model uses different k-distances, resulting in spending additional time on performing calculations. We calculate the excessive wall time as the difference between the wall time corresponding to k-distance used by ML-model/baseline model and the mean optimal wall time in case when predicted k-distance is smaller or equal than the true one (converged calculation), and in opposite case all computation time is wasted, so all the time predicted by the model is added to the excessive time:
\begin{equation}
T_{\mathrm{exc}} =
\begin{cases}
T_{\mathrm{model}} - T_{\mathrm{true}}, & k_{\mathrm{dist}}^{\mathrm{pred}} \le k_{\mathrm{dist}}^{\mathrm{true}}, \\[6pt]
T_{\mathrm{model}}, & k_{\mathrm{dist}}^{\mathrm{pred}} > k_{\mathrm{dist}}^{\mathrm{true}} .
\end{cases}
\end{equation}
Interestingly, the gain in computation time for ML prediction compared to the baseline model depends on the quantile. For higher quantiles and more accurate calculations, the gain is more noticeable. With a decrease in quantile, the number of mispredicted samples grows, and the gain in time for ML predictions compared to bthe aseline model disappears. 
\begin{table*}
  \caption{Wall-time analysis for predicted $k_{\mathrm{dist}}$ compared with optimal and dummy baselines. 
  Excessive wall time is defined relative to the wall time at optimal $k_{\mathrm{dist}}$.}
  \label{tbl:walltime-analysis}
  \setlength{\tabcolsep}{3pt}
   \begin{ruledtabular}
    \scriptsize
  \begin{tabular*}{\textwidth}{@{\extracolsep{\fill}}lllllllll}
    \# & Model & Features & \makecell{Mean\\wall time\\
    (optimal), s} & \makecell{Confidence\\level} &
        \makecell{Excessive\\wall time\\(pred.), s} & \makecell{Excessive\\ wall time\\(dummy), s} &
        \makecell{Gain\\per sample, s} & Gain, \% \\
    \hline
    1 & RF      & CSLM                     & 352 & 0.95 & 388 & 497 & 109 & 22 \\
    2 & RF      & CSLM                     & 352 & 0.90 & 277 & 318 & 41  & 13 \\
    3 & RF      & CSLM                     & 352 & 0.85 & 240 & 223 & 17  & 7  \\
    4 & ALIGNNd & rg+cgcnn+edc+CSL         & 352 & 0.95 & 368 & 468 & 100 & 21 \\
    5 & ALIGNNd & rg+cgcnn+edc+CSL         & 352 & 0.90 & 247 & 311 & 64  & 21 \\
    6 & ALIGNNd & rg+cgcnn+edc+CSL         & 352 & 0.85 & 217 & 217 & 0   & 0  \\
  \end{tabular*}
  \end{ruledtabular}
\end{table*}

\section{Application for input file generation}
To share the models we optimized, we developed a web application that allows to generation of an input file for Quantum Espresso single-point SCF total energy calculations, https://goldilocks.streamlit.app/. 

The input page of the application asks the user to choose the variant of SSSP1.3 PBEsol to use, efficiency or precision, the ML model, and the confidence level. Our models are trained on data generated with SSSP1.3 PBEsol efficiency, but as shown above, for a large fraction of compounds, the k-points convergence values remain the same for SSSP1.3 PBEsol precision.  

Next, the user is prompted to upload the structure CIF file or query the structure from one of the databases. Then, the structure may be subjected to transformations: reduced to a primitive one, or multiplied to create a super cell. When all parameters are specified, the user is prompted to generate the input file. Input file generation can be made with the ASE deterministic function, or the LLM (the choice is available). The file generator makes prediction of the k-point distance using a primitive cell of the compound (as all structures in the training dataset were pre-processed into primitive ones), makes prediction of k-point density, and then calculates k-mesh from the inverse unit cell and predicted k-point distance.  

\section{Discussion}
The results presented above show that machine-learning models, when trained on a sufficiently large and diverse dataset of converged calculations, can reliably predict k-point densities for plane-wave DFT simulations. More importantly, they can do so with statistical guarantees when combined with conformalised quantile regression. This is particularly relevant for high-throughput and agentic workflows, where parameter selection is a major bottleneck.

However, several limitations remain. First, magnetic materials remain challenging: spin-polarized calculations often yield different k-mesh requirements, and our current dataset includes only non-spin-polarized data. Second, the dataset is tied to SSSP-1.3 PBEsol pseudopotentials; although we show that trends transfer to the Precision library, other pseudopotential families may require retraining. Third, while our models capture structural and compositional factors well, more expressive representations—such as equivariant GNNs or learned Brillouin-zone features—may further improve accuracy. Next, we did not try to pre-train GNN/CrabNet models on the metallicity dataset. Transfer learning could work better than feature injection. Finally, we could use a completely different approach to building an ML model: we could predict total energy conditioned on structure and other DFT parameters. Then the speed up of the solution of the convergence problem would be realized through using an ML surrogate to generate the convergence curve instead of predicting the converged k-distance. If successful, this approach could be applicable more generally. However, it is not immediately clear how to determine convergence in this case, as ML predictions of the total energy are expected to introduce errors, which could obscure the convergence point. Maybe a combination of both approaches could be used.

Despite these limitations, the current approach is successful in reaching the objective to improve over the constant baseline, which is currently is a common choice. The integration with MCP servers also opens a path toward fully autonomous computational agents capable of end-to-end materials simulation.

\section*{Conclusions}
In summary, we have developed a machine-learning framework that predicts the k-point density required for converged Quantum ESPRESSO SCF energy calculations. Using a dataset of more than 20,000 convergence-tested materials, we benchmarked a broad set of models and demonstrated that both ensemble methods and graph neural networks achieve high predictive accuracy. Combining these models with conformalised quantile regression enables statistically guaranteed, underestimation-averse predictions.

Relative to fixed-density workflows, the model reduces unnecessary k-mesh sampling and thereby lowers computational cost while maintaining accuracy. The provided web platform makes this tool available. We also plan to release an MCP server in the near future to make our tool available in any MCP-compatible system.

Overall, this work establishes a practical and scalable route to data-driven parameter optimisation in DFT, opening doors to more efficient, greener, and more robust high-throughput materials simulations.

\section*{Author contributions}
Elena Patyukova: conceptualization (AI), methodology (AI), investigation, formal analysis, software, validation, writing – original draft, visualization. Junwen Yin: conceptualization (DFT), methodology (DFT), data curation, investigation, writing – original draft, visualization. Susmita Basak: investigation, funding acquisition. Samuel Pinilla Sanchez: investigation. Alin Elena: data curation, resources, funding acquisition. Gilberto Teobaldi: writing – review \& editing, project administration, supervision. 

\section*{Conflicts of interest}
There are no conflicts to declare.

\section*{Data availability}
The generated data is available under CC-BY-4.0 license, the repository is located at \href{https://data-collections.psdi.ac.uk/records/75959-bwa52}{https://data-collections.psdi.ac.uk/records/75959-bwa52}. The code for the web application can be found at \href{https://github.com/stfc/goldilocks}{https://github.com/stfc/goldilocks}. Code used for training the models can be found at\href{https://github.com/stfc/goldilocks_kpoints}{https://github.com/stfc/goldilocks\_kpoints}. The web application can be found at  https://goldilocks.streamlit.app/.

\section*{Acknowledgements}

This work was carried out thanks to EPSRC funding awarded to Barbara Montanari, Alin Elena, and Susmita Basak: Goldilocks convergence tools and best practices for numerical approximations in Density Functional Theory calculations (EP/Z530657/1). This work was co-funded by the Ada Lovelace Centre at STFC-UKRI. Computing resources were provided by STFC Scientific Computing Department’s SCARF cluster and STFC cloud.

\bibliography{apssamp}

@article{Giannozzi_2009,
doi = {10.1088/0953-8984/21/39/395502},
url = {https://doi.org/10.1088/0953-8984/21/39/395502},
year = {2009},
month = {sep},
publisher = {},
volume = {21},
number = {39},
pages = {395502},
author = {Giannozzi, Paolo and Baroni, Stefano and Bonini, Nicola and Calandra, Matteo and Car, Roberto and Cavazzoni, Carlo and Ceresoli, Davide and Chiarotti, Guido L and Cococcioni, Matteo and Dabo, Ismaila and Dal Corso, Andrea and de Gironcoli, Stefano and Fabris, Stefano and Fratesi, Guido and Gebauer, Ralph and Gerstmann, Uwe and Gougoussis, Christos and Kokalj, Anton and Lazzeri, Michele and Martin-Samos, Layla and Marzari, Nicola and Mauri, Francesco and Mazzarello, Riccardo and Paolini, Stefano and Pasquarello, Alfredo and Paulatto, Lorenzo and Sbraccia, Carlo and Scandolo, Sandro and Sclauzero, Gabriele and Seitsonen, Ari P and Smogunov, Alexander and Umari, Paolo and Wentzcovitch, Renata M},
title = {QUANTUM ESPRESSO: a modular and open-source software project for quantum
simulations of materials},
journal = {Journal of Physics: Condensed Matter}
}

@article{Huber2020,
  author    = {Huber, Sebastiaan P. and Zoupanos, Spyros and Uhrin, Martin and Talirz, Leopold and Kahle, Leonid and H{\"a}uselmann, Rico and Gresch, Dominik and M{\"u}ller, Tiziano and Yakutovich, Aliaksandr V. and Andersen, Casper W. and Ramirez, Francisco F. and Adorf, Carl S. and Gargiulo, Fernando and Kumbhar, Snehal and Passaro, Elsa and Johnston, Conrad and Merkys, Andrius and Cepellotti, Andrea and Mounet, Nicolas and Marzari, Nicola and Kozinsky, Boris and Pizzi, Giovanni},
  title     = {AiiDA 1.0, a scalable computational infrastructure for automated reproducible workflows and data provenance},
  journal   = {Scientific Data},
  volume    = {7},
  number    = {1},
  pages     = {300},
  year      = {2020},
  doi       = {10.1038/s41597-020-00638-4},
  url       = {https://doi.org/10.1038/s41597-020-00638-4},
  issn      = {2052-4463},
}

@article{Ganose2025,
author ="Ganose, Alex M. and Sahasrabuddhe, Hrushikesh and Asta, Mark and Beck, Kevin and Biswas, Tathagata and Bonkowski, Alexander and Bustamante, Joana and Chen, Xin and Chiang, Yuan and Chrzan, Daryl C. and Clary, Jacob and Cohen, Orion A. and Ertural, Christina and Gallant, Max C. and George, Janine and Gerits, Sophie and Goodall, Rhys E. A. and Guha, Rishabh D. and Hautier, Geoffroy and Horton, Matthew and Inizan, T. J. and Kaplan, Aaron D. and Kingsbury, Ryan S. and Kuner, Matthew C. and Li, Bryant and Linn, Xavier and McDermott, Matthew J. and Mohanakrishnan, Rohith Srinivaas and Naik, Aakash N. and Neaton, Jeffrey B. and Parmar, Shehan M. and Persson, Kristin A. and Petretto, Guido and Purcell, Thomas A. R. and Ricci, Francesco and Rich, Benjamin and Riebesell, Janosh and Rignanese, Gian-Marco and Rosen, Andrew S. and Scheffler, Matthias and Schmidt, Jonathan and Shen, Jimmy-Xuan and Sobolev, Andrei and Sundararaman, Ravishankar and Tezak, Cooper and Trinquet, Victor and Varley, Joel B. and Vigil-Fowler, Derek and Wang, Duo and Waroquiers, David and Wen, Mingjian and Yang, Han and Zheng, Hui and Zheng, Jiongzhi and Zhu, Zhuoying and Jain, Anubhav",
title  ="Atomate2: modular workflows for materials science",
journal  ="Digital Discovery",
year  ="2025",
volume  ="4",
issue  ="7",
pages  ="1944-1973",
publisher  ="RSC",
doi  ="10.1039/D5DD00019J",
url  ="http://dx.doi.org/10.1039/D5DD00019J"}

@article{JANSSEN2019,
title = {pyiron: An integrated development environment for computational materials science},
journal = {Computational Materials Science},
volume = {163},
pages = {24-36},
year = {2019},
issn = {0927-0256},
doi = {https://doi.org/10.1016/j.commatsci.2018.07.043},
url = {https://www.sciencedirect.com/science/article/pii/S0927025618304786},
author = {Jan Janssen and Sudarsan Surendralal and Yury Lysogorskiy and Mira Todorova and Tilmann Hickel and Ralf Drautz and Jörg Neugebauer}
}

@article{Curtarolo2012,
title = {AFLOW: An automatic framework for high-throughput materials discovery},
journal = {Computational Materials Science},
volume = {58},
pages = {218-226},
year = {2012},
issn = {0927-0256},
doi = {https://doi.org/10.1016/j.commatsci.2012.02.005},
url = {https://www.sciencedirect.com/science/article/pii/S0927025612000717},
author = {Stefano Curtarolo and Wahyu Setyawan and Gus L.W. Hart and Michal Jahnatek and Roman V. Chepulskii and Richard H. Taylor and Shidong Wang and Junkai Xue and Kesong Yang and Ohad Levy and Michael J. Mehl and Harold T. Stokes and Denis O. Demchenko and Dane Morgan}
}

@article{Zou2025,
   title={El Agente: An autonomous agent for quantum chemistry},
   volume={8},
   ISSN={2590-2385},
   url={http://dx.doi.org/10.1016/j.matt.2025.102263},
   DOI={10.1016/j.matt.2025.102263},
   number={7},
   journal={Matter},
   publisher={Elsevier BV},
   author={Zou, Yunheng and Cheng, Austin H. and Aldossary, Abdulrahman and Bai, Jiaru and Leong, Shi Xuan and Campos-Gonzalez-Angulo, Jorge Arturo and Choi, Changhyeok and Ser, Cher Tian and Tom, Gary and Wang, Andrew and Zhang, Zijian and Yakavets, Ilya and Hao, Han and Crebolder, Chris and Bernales, Varinia and Aspuru-Guzik, Alán},
   year={2025},
   month=jul, pages={102263} }

@misc{wang2025dreamsdensityfunctionaltheory,
      title={DREAMS: Density Functional Theory Based Research Engine for Agentic Materials Simulation}, 
      author={Ziqi Wang and Hongshuo Huang and Hancheng Zhao and Changwen Xu and Shang Zhu and Jan Janssen and Venkatasubramanian Viswanathan},
      year={2025},
      eprint={2507.14267},
      archivePrefix={arXiv},
      primaryClass={cs.AI},
      url={https://arxiv.org/abs/2507.14267}, 
}

@article{Liu2025,
doi = {10.1088/1674-1056/ae0681},
url = {https://doi.org/10.1088/1674-1056/ae0681},
year = {2025},
month = {nov},
publisher = {Chinese Physical Society and IOP Publishing Ltd},
volume = {34},
number = {11},
pages = {117106},
author = {Liu, Jiaxuan and Zhu, Tiannian and Ye, Caiyuan and Fang, Zhong and Weng, Hongming and Wu, Quansheng},
title = {VASPilot: MCP-facilitated multi-agent intelligence for autonomous VASP simulations},
journal = {Chinese Physics B}
}

@article{Choudhary2019,
title = {Convergence and machine learning predictions of Monkhorst-Pack k-points and plane-wave cut-off in high-throughput DFT calculations},
journal = {Computational Materials Science},
volume = {161},
pages = {300-308},
year = {2019},
issn = {0927-0256},
doi = {https://doi.org/10.1016/j.commatsci.2019.02.006},
url = {https://www.sciencedirect.com/science/article/pii/S0927025619300813},
author = {Kamal Choudhary and Francesca Tavazza}
}

@misc{janssen2021automatedoptimizationconvergenceparameters,
      title={Automated optimization of convergence parameters in plane wave density functional theory calculations via a tensor decomposition-based uncertainty quantification}, 
      author={Jan Janssen and Edgar Makarov and Tilmann Hickel and Alexander V. Shapeev and Jörg Neugebauer},
      year={2021},
      eprint={2112.04081},
      archivePrefix={arXiv},
      primaryClass={cond-mat.mtrl-sci},
      url={https://arxiv.org/abs/2112.04081}, 
}

@article{Prandini2018,
  author    = {Prandini, Gianluca and Marrazzo, Antimo and Castelli, Ivano E. and Mounet, Nicolas and Marzari, Nicola},
  title     = {Precision and efficiency in solid-state pseudopotential calculations},
  journal   = {npj Computational Materials},
  volume    = {4},
  number    = {1},
  pages     = {72},
  year      = {2018},
  doi       = {10.1038/s41524-018-0127-2},
  url       = {https://doi.org/10.1038/s41524-018-0127-2},
  issn      = {2057-3960}
}

@article{Bosoni2024,
  author    = {Bosoni, Emanuele and Beal, Louis and Bercx, Marnik and Blaha, Peter and Bl{\"u}gel, Stefan and Br{\"o}der, Jens and Callsen, Martin and Cottenier, Stefaan and Degomme, Augustin and Dikan, Vladimir and Eimre, Kristjan and Flage-Larsen, Espen and Fornari, Marco and Garcia, Alberto and Genovese, Luigi and Giantomassi, Matteo and Huber, Sebastiaan P. and Janssen, Henning and Kastlunger, Georg and Krack, Matthias and Kresse, Georg and K{\"u}hne, Thomas D. and Lejaeghere, Kurt and Madsen, Georg K. H. and Marsman, Martijn and Marzari, Nicola and Michalicek, Gregor and Mirhosseini, Hossein and M{\"u}ller, Tiziano M. A. and Petretto, Guido and Pickard, Chris J. and Ponc{\'e}, Samuel and Rignanese, Gian-Marco and Rubel, Oleg and Ruh, Thomas and Sluydts, Michael and Vanpoucke, Danny E. P. and Vijay, Sudarshan and Wolloch, Michael and Wortmann, Daniel and Yakutovich, Aliaksandr V. and Yu, Jusong and Zadoks, Austin and Zhu, Bonan and Pizzi, Giovanni},
  title     = {How to verify the precision of density-functional-theory implementations via reproducible and universal workflows},
  journal   = {Nature Reviews Physics},
  volume    = {6},
  number    = {1},
  pages     = {45--58},
  year      = {2024},
  doi       = {10.1038/s42254-023-00655-3},
  url       = {https://doi.org/10.1038/s42254-023-00655-3},
  issn      = {2522-5820}
}

@misc{nascimento2025accurateefficientprotocolshighthroughput,
      title={Accurate and efficient protocols for high-throughput first-principles materials simulations}, 
      author={Gabriel de Miranda Nascimento and Flaviano José dos Santos and Marnik Bercx and Davide Grassano and Giovanni Pizzi and Nicola Marzari},
      year={2025},
      eprint={2504.03962},
      archivePrefix={arXiv},
      primaryClass={cond-mat.mtrl-sci},
      url={https://arxiv.org/abs/2504.03962}, 
}

@article{Ward2016,
  author    = {Ward, Logan and Agrawal, Ankit and Choudhary, Alok and Wolverton, Christopher},
  title     = {A general-purpose machine learning framework for predicting properties of inorganic materials},
  journal   = {npj Computational Materials},
  volume    = {2},
  number    = {1},
  pages     = {16028},
  year      = {2016},
  doi       = {10.1038/npjcompumats.2016.28},
  url       = {https://doi.org/10.1038/npjcompumats.2016.28},
  issn      = {2057-3960}
}

@article{Wang2021,
  author    = {Wang, Anthony Yu-Tung and Kauwe, Steven K. and Murdock, Ryan J. and Sparks, Taylor D.},
  title     = {Compositionally restricted attention-based network for materials property predictions},
  journal   = {npj Computational Materials},
  volume    = {7},
  number    = {1},
  pages     = {77},
  year      = {2021},
  doi       = {10.1038/s41524-021-00545-1},
  url       = {https://doi.org/10.1038/s41524-021-00545-1},
  issn      = {2057-3960}
}

@article{Xie2018,
  title = {Crystal Graph Convolutional Neural Networks for an Accurate and Interpretable Prediction of Material Properties},
  author = {Xie, Tian and Grossman, Jeffrey C.},
  journal = {Phys. Rev. Lett.},
  volume = {120},
  issue = {14},
  pages = {145301},
  numpages = {6},
  year = {2018},
  month = {Apr},
  publisher = {American Physical Society},
  doi = {10.1103/PhysRevLett.120.145301},
  url = {https://link.aps.org/doi/10.1103/PhysRevLett.120.145301}
}

@article{Choudhary2021,
  author    = {Choudhary, Kamal and DeCost, Brian},
  title     = {Atomistic Line Graph Neural Network for improved materials property predictions},
  journal   = {npj Computational Materials},
  volume    = {7},
  number    = {1},
  pages     = {185},
  year      = {2021},
  doi       = {10.1038/s41524-021-00650-1},
  url       = {https://doi.org/10.1038/s41524-021-00650-1},
  issn      = {2057-3960}
}

@article{
Sandip2016,
author ="De, Sandip and Bartók, Albert P. and Csányi, Gábor and Ceriotti, Michele",
title  ="Comparing molecules and solids across structural and alchemical space",
journal  ="Phys. Chem. Chem. Phys.",
year  ="2016",
volume  ="18",
issue  ="20",
pages  ="13754-13769",
publisher  ="The Royal Society of Chemistry",
doi  ="10.1039/C6CP00415F",
url  ="http://dx.doi.org/10.1039/C6CP00415F"}

@article{Himanen2020,
title = {DScribe: Library of descriptors for machine learning in materials science},
journal = {Computer Physics Communications},
volume = {247},
pages = {106949},
year = {2020},
issn = {0010-4655},
doi = {https://doi.org/10.1016/j.cpc.2019.106949},
url = {https://www.sciencedirect.com/science/article/pii/S0010465519303042},
author = {Lauri Himanen and Marc O.J. Jäger and Eiaki V. Morooka and Filippo {Federici Canova} and Yashasvi S. Ranawat and David Z. Gao and Patrick Rinke and Adam S. Foster}
}

@article{Tshitoyan2019,
  author    = {Tshitoyan, Vahe and Dagdelen, John and Weston, Leigh and Dunn, Alexander and Rong, Ziqin and Kononova, Olga and Persson, Kristin A. and Ceder, Gerbrand and Jain, Anubhav},
  title     = {Unsupervised word embeddings capture latent knowledge from materials science literature},
  journal   = {Nature},
  volume    = {571},
  number    = {7763},
  pages     = {95--98},
  year      = {2019},
  doi       = {10.1038/s41586-019-1335-8},
  url       = {https://doi.org/10.1038/s41586-019-1335-8},
  issn      = {1476-4687}
}

@article{Gong2023,
author = {Sheng Gong  and Keqiang Yan  and Tian Xie  and Yang Shao-Horn  and Rafael Gomez-Bombarelli  and Shuiwang Ji  and Jeffrey C. Grossman },
title = {Examining graph neural networks for crystal structures: Limitations and opportunities for capturing periodicity},
journal = {Science Advances},
volume = {9},
number = {45},
pages = {eadi3245},
year = {2023},
doi = {10.1126/sciadv.adi3245},
URL = {https://www.science.org/doi/abs/10.1126/sciadv.adi3245},
eprint = {https://www.science.org/doi/pdf/10.1126/sciadv.adi3245}}

@article{Meinshausen2006,
  author  = {Nicolai Meinshausen},
  title   = {Quantile Regression Forests},
  journal = {Journal of Machine Learning Research},
  year    = {2006},
  volume  = {7},
  number  = {35},
  pages   = {983--999},
  url     = {http://jmlr.org/papers/v7/meinshausen06a.html}
}

@misc{romano2019conformalizedquantileregression,
      title={Conformalized Quantile Regression}, 
      author={Yaniv Romano and Evan Patterson and Emmanuel J. Candès},
      year={2019},
      eprint={1905.03222},
      archivePrefix={arXiv},
      primaryClass={stat.ME},
      url={https://arxiv.org/abs/1905.03222}, 
}

@article{Uhrin2021,
title = {Workflows in AiiDA: Engineering a high-throughput, event-based engine for robust and modular computational workflows},
journal = {Computational Materials Science},
volume = {187},
pages = {110086},
year = {2021},
issn = {0927-0256},
doi = {https://doi.org/10.1016/j.commatsci.2020.110086},
url = {https://www.sciencedirect.com/science/article/pii/S0927025620305772},
author = {Martin Uhrin and Sebastiaan P. Huber and Jusong Yu and Nicola Marzari and Giovanni Pizzi}
}

@article{Ong2013,
title = {Python Materials Genomics (pymatgen): A robust, open-source python library for materials analysis},
journal = {Computational Materials Science},
volume = {68},
pages = {314-319},
year = {2013},
issn = {0927-0256},
doi = {https://doi.org/10.1016/j.commatsci.2012.10.028},
url = {https://www.sciencedirect.com/science/article/pii/S0927025612006295},
author = {Shyue Ping Ong and William Davidson Richards and Anubhav Jain and Geoffroy Hautier and Michael Kocher and Shreyas Cholia and Dan Gunter and Vincent L. Chevrier and Kristin A. Persson and Gerbrand Ceder}
}

@misc{huber2025mc3dmaterialscloudcomputational,
      title={MC3D: The Materials Cloud computational database of experimentally known stoichiometric inorganics}, 
      author={Sebastiaan P. Huber and Michail Minotakis and Marnik Bercx and Timo Reents and Kristjan Eimre and Nataliya Paulish and Nicolas Hörmann and Martin Uhrin and Nicola Marzari and Giovanni Pizzi},
      year={2025},
      eprint={2508.19223},
      archivePrefix={arXiv},
      primaryClass={cond-mat.mtrl-sci},
      url={https://arxiv.org/abs/2508.19223}, 
}

@article{Jain2013,
    author = {Jain, Anubhav and Ong, Shyue Ping and Hautier, Geoffroy and Chen, Wei and Richards, William Davidson and Dacek, Stephen and Cholia, Shreyas and Gunter, Dan and Skinner, David and Ceder, Gerbrand and Persson, Kristin A.},
    title = {Commentary: The Materials Project: A materials genome approach to accelerating materials innovation},
    journal = {APL Materials},
    volume = {1},
    number = {1},
    pages = {011002},
    year = {2013},
    month = {07},
    issn = {2166-532X},
    doi = {10.1063/1.4812323},
    url = {https://doi.org/10.1063/1.4812323},
}

@article{Breiman2001,
  author    = {Breiman, Leo},
  title     = {Random Forests},
  journal   = {Machine Learning},
  volume    = {45},
  number    = {1},
  pages     = {5--32},
  year      = {2001},
  doi       = {10.1023/A:1010933404324},
  url       = {https://doi.org/10.1023/A:1010933404324},
  issn      = {1573-0565}
}

@article{Friedman2001,
  author    = {Friedman, Jerome H.},
  title     = {Greedy Function Approximation: A Gradient Boosting Machine},
  journal   = {The Annals of Statistics},
  volume    = {29},
  number    = {5},
  pages     = {1189--1232},
  year      = {2001},
  url       = {http://www.jstor.org/stable/2699986}
}

@article{Ward2018,
title = {Matminer: An open source toolkit for materials data mining},
journal = {Computational Materials Science},
volume = {152},
pages = {60-69},
year = {2018},
issn = {0927-0256},
doi = {https://doi.org/10.1016/j.commatsci.2018.05.018},
url = {https://www.sciencedirect.com/science/article/pii/S0927025618303252},
author = {Logan Ward and Alexander Dunn and Alireza Faghaninia and Nils E.R. Zimmermann and Saurabh Bajaj and Qi Wang and Joseph Montoya and Jiming Chen and Kyle Bystrom and Maxwell Dylla and Kyle Chard and Mark Asta and Kristin A. Persson and G. Jeffrey Snyder and Ian Foster and Anubhav Jain}
}

@misc{vaswani2023attentionneed,
      title={Attention Is All You Need}, 
      author={Ashish Vaswani and Noam Shazeer and Niki Parmar and Jakob Uszkoreit and Llion Jones and Aidan N. Gomez and Lukasz Kaiser and Illia Polosukhin},
      year={2023},
      eprint={1706.03762},
      archivePrefix={arXiv},
      primaryClass={cs.CL},
      url={https://arxiv.org/abs/1706.03762}, 
}

@book{molnar2025,
  author    = {Molnar, Christoph},
  title     = {Interpretable Machine Learning},
  subtitle  = {A Guide for Making Black Box Models Explainable},
  edition   = {3},
  year      = {2025},
  isbn      = {978-3-911578-03-5},
  url       = {https://christophm.github.io/interpretable-ml-book}
}

\end{document}